\documentclass[
 reprint,
superscriptaddress,
 amsmath,amssymb,
 aps,
]{revtex4-2}

\usepackage{graphicx}
\usepackage{dcolumn}
\usepackage{bm}
\usepackage{xcolor}

\begin{document}

\preprint{APS/123-QED}

\title{Intrinsic electronic phase separation and competition between $G$-type, $C$-type and $CE$-type charge and orbital ordering modes in Hg$_{1-x}$Na$_x$Mn$_3$Mn$_4$O$_{12}$} 

\author{Ben R. M. Tragheim}
\affiliation{Department of Chemistry, University of Warwick, Gibbet Hill, Coventry, CV4 7AL, UK}

\author{Struan Simpson}
\affiliation{Department of Chemistry, University of Warwick, Gibbet Hill, Coventry, CV4 7AL, UK}

\author{En-Pei Liu}
\affiliation{Center for Condensed Matter Sciences, National Taiwan University, Taipei 10617, Taiwan}
\affiliation{Department of Physics, National Taiwan University, Taipei 10617, Taiwan}

\author{Mark S. Senn}
\email{m.senn@warwick.ac.uk}
\affiliation{Department of Chemistry, University of Warwick, Gibbet Hill, Coventry, CV4 7AL, UK}

\author{Wei-Tin Chen}
\email{weitinchen@ntu.edu.tw}
\affiliation{Center for Condensed Matter Sciences, National Taiwan University, Taipei 10617, Taiwan}
\affiliation{Center of Atomic Initiative for New Materials, National Taiwan University, Taipei 10617, Taiwan}
\affiliation{Taiwan Consortium of Emergent Crystalline Materials, National Science and Technology Council, Taipei 10622, Taiwan}

\date{\today}

\begin{abstract}

The novel series of hole-doped quadruple manganite perovskites Hg$_{1-x}$Na$_x$Mn$_3$Mn$_4$O$_{12}$ (HNMO) has been synthesized and its charge and orbital order behavior investigated through high-resolution synchrotron powder x-ray diffraction techniques. Through careful Rietveld refinements of structural models $via$ symmetry-motivated approaches, we show that the ground state of HNMO compositions adopts a polar $G$-type charge and orbital ordered state, which is rare in manganite perovskites, and is robust as a sole phase up to a critical doping level. Upon this critical doping, coincident with that in which colossal magnetoresistance (CMR) is maximal in canonical manganite perovskites, electronic phase separation occurs between $G$-type and orbital order with charge disorder-type states. The latter state has recently been identified in Ca$_{1-x}$Na$_x$Mn$_3$Mn$_4$O$_{12}$ perovskites, and proposed to be the competing insulating state from which CMR phenomena emerges. We show the mechanism for the formation of the $G$-type state is due to charge transfer processes which may occur through a coupling of distortions involving structural and charge and orbital degrees of freedom, ultimately driving the polar ground state through an improper-like ferroelectric polarization mechanism. These results will act as an important recipe for designing novel ferroelectric-active materials, in addition to expanding the richness of charge and orbital ordered states in manganite perovskites.     

\end{abstract}

\maketitle


\section{\label{sec:level1}Introduction}

\begin{figure}[!t]
\includegraphics{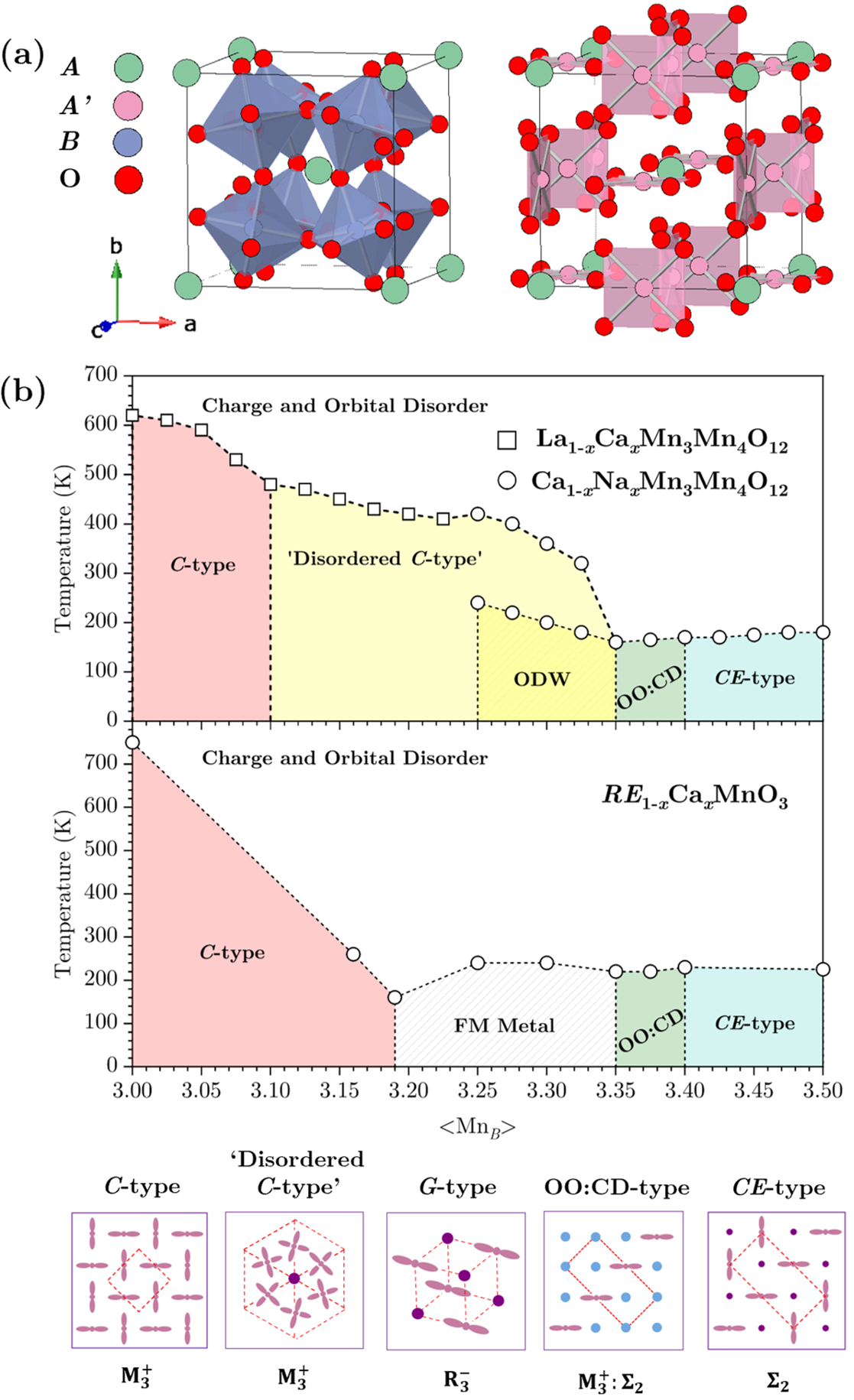}
\caption{\label{quadruple-perovskite-ordering-schemes} (a) Schematic of the $A$Mn$_3$Mn$_4$O$_{12}$ quadruple perovskite structure with key structural motifs $B$O$_6$ octahedra and $A'$O$_4$ square planar units shown separately for clarity. (b) Charge and orbital order phase diagrams for quadruple perovskites La$_{1-x}$Ca$_x$Mn$_3$Mn$_4$O$_{12}$, Ca$_{1-x}$Na$_x$Mn$_3$Mn$_4$O$_{12}$, and general $RE_{1-x}$Ca$_x$MnO$_3$ perovskites. Data points for the quadruple perovskites are taken from our previous study.\cite{chen2021striping} Data points for the $RE_{1-x}$Ca$_x$MnO$_3$ perovskites are taken from various LCMO and LPCMO compositions.\cite{radaelli1995simultaneous, hwang1995lattice, radaelli1997charge, rodriguez1998neutron, van2003orbital, chen2021striping} $<$Mn$_B$$>$ is given with respect to the high-temperature state of each system before any charge ordering occurs. ODW $=$ orbital density wave. FM $=$ ferromagnetic. Below are representations of the different types of charge and orbital ordered states discussed in the current work. Pink lobes, purple dots and blue dots correspond to Mn$^{3+}$, Mn$^{4+}$ and Mn$^{3.5+}$ respectively residing on the perovskite $B$ site. Red dashed boxes indicate the minimum repeating unit required to describe each state. Structural distortions resulting in each state transform as the irrep given below each ordered state. $C$-, OO:CD- and $CE$-type states show projections in the $ab$ crystallographic plane, `disordered $C$-type' and $G$-type states are shown along the [111] direction of the quadruple perovskite structure.}
\end{figure}

Manganite perovskites of the structural form $A$MnO$_3$ have been intensely studied for their intricate interplay between structural, orbital, electronic and magnetic degrees of freedom. Key consequences of this interplay are the formation of unique charge and orbital (dis)ordered states, including $C$-type orbital order in LaMnO$_3$ and $CE$-type charge and orbital order in La$_{0.5}$Ca$_{0.5}$MnO$_3$ (Figure \ref{quadruple-perovskite-ordering-schemes}).\cite{wollan1955neutron, goodenough1955theory, radaelli1997charge} Each of these states are crucial in the manifestation of many properties such as metal-to-insulator transitions, (anti)ferromagnetism and colossal magnetoresistance (CMR). Hence, studying the nature of these charge and orbital-based states is paramount to establish structure-properties relationships, aiding in the design and optimization of novel functional materials.
 
Prominent derivatives of the $A$MnO$_3$ perovskite, the quadruple perovskites $A$Mn$^{A'}_3$Mn$^B_4$O$_{12}$ (Figure \ref{quadruple-perovskite-ordering-schemes}), serve to expand the richness of different unique charge and orbital (dis)ordered states. These are related to the aristotypical $AB$O$_3$ perovskite structure where they contain four equivalents of the $AB$O$_3$ structural unit, but with these $A$ sites split into a 1:3 ratio of $A$:$A'$ cations. $A$ cations contain the nominal 12-fold coordination environment as in $AB$O$_3$ perovskites, but the $A'$ site exhibits a square planar configuration and is occupied by Mn$^{3+}$. The Jahn-Teller (JT)-active cation on these $A$' sites results in quadruple perovskites containing a much larger degree of $B$O$_6$ octahedral rotations compared to their $AB$O$_3$ counterparts, whilst also `locking in' their magnitude by restricting their ability to undergo additional rotations. An immediate example of the expanded ordering states observable in these quadruple perovskites are the intermediate temperature $R\bar{3}$ phases of divalent $A$Mn$_3$Mn$_4$O$_{12}$, $A$ $=$ Ca, Sr, Cd, Pb,\cite{przenioslo2002phase, glazkova2015high, locherer2012synthesis} perovskites exhibiting a `disordered $C$-type' state (Figure \ref{quadruple-perovskite-ordering-schemes}). This is best described as a charge order of Mn$^{4+}$ on $1/4$ of the $B$ sites along the [111] direction of the $R\bar{3}$ unit cell (in the rhombohedral setting), with a disordering of $d_{z^2}$-type orbitals of Mn$^{3+}$ on every other $B$ site. The orbital disorder is attributed from the formation of MnO$_6$ octahedra with an unusual `4-long 2-short' configuration of Mn-O bond lengths, which due to higher-order anharmonic JT effects is energetically unfavourable to occur compared to the usual `2-long 4-short' configuration.\cite{khomskii2000anharmonic} In terms of their low temperature ground states these divalent manganite quadruple perovskites are known to exhibit incommensurate modulations of their crystal and magnetic structures,\cite{perks2012magneto, belik2016low, johnson2017magneto, guo2017non, johnson2021competing} giving rise to exotic magneto-orbital textures and the formation of orbital density waves.\cite{perks2012magneto} Hence, these material systems prove essential to be studied for both fundamental and application-driven motivations based on their structural, orbital and magnetic behaviors. 

Systematic substitution of the divalent $A$ site in CaMn$_3$Mn$_4$O$_{12}$ for the monovalent cation Na$^+$ results in a gradual hole doping of the Mn $B$ site whilst preserving the oxidation state of the Mn$^{3+}$ $A'$ site. In this case the average oxidation state of the $B$ site, $<$Mn$_B$$>$, varies systematically from $+$3.25 to $+$3.5 through the solid solution Ca$_{1-x}$Na$_x$Mn$_3$Mn$_4$O$_{12}$. Here, $<$Mn$_B$$>$ is derived from the high-temperature $Im\bar{3}$ phase before any kind of charge ordering occurs. We have previously studied this solid solution and we discovered at optimal doping regimes producing $<$Mn$_B$$>$ $\approx$ $+$3.375 ($3/8^{th}$ doping level, corresponding to $x$ $\approx$ 0.4/0.5) the formation of the novel orbital order:charge disorder (OO:CD)-type state arises, as shown in Figure \ref{quadruple-perovskite-ordering-schemes}.\cite{chen2021striping} This optimal doping coincides exactly with $<$Mn$_B$$>$ that occurs in CMR-active manganites La$_{1-x}$Ca$_x$MnO$_3$ (LCMO) and La$_{1-x-y}$Pr$_y$Ca$_x$MnO$_3$ (LPCMO) where the CMR response is at its maximum.\cite{cheong2000colossal, hwang1995lattice, hwang1995pressure} The OO:CD-type state is described as containing a single diagonal stripe of orbital order due to Mn$^{3+}$ which repeats every fourth stripe, with a charge disorder reservoir of Mn$^{3.5+}$ between these stripes. The mechanism through which this novel state arises is due to the equal ratio of structural distortion modes that, by themselves, constitute pure $C$-type orbital order and pure $CE$-type charge and orbital order, and couple to the emergence of a pseudo-tetragonal lattice strain state.

Another novel ordering state in the manganite perovskites is the $G$-type charge and orbital order state (Figure \ref{quadruple-perovskite-ordering-schemes}) in the system HgMn$_3$Mn$_4$O$_{12}$ (HMO) which is in principle isoelectronic with CaMn$_3$Mn$_4$O$_{12}$.\cite{chen2018improper} Like other divalent manganite quadruple perovskites, HMO undergoes the series of structural phase transitions $Im\bar{3}$ to $R\bar{3}$ where the $R\bar{3}$ phase demonstrates the `disordered $C$-type' state. However, unique to HMO a successive phase transition from $R\bar{3}$ to $P$2 ($Pnn$2 pseudo-symmetry) was observed to occur at $T$ $=$ 240 K. One of the key structural distortions in this novel state transforms as the irreducible representation (irrep) R$_3^-$ (with respect to the $Pm\bar{3}m$ $AB$O$_3$ perovskite structure, $A$ site origin) which gives rise to the $G$-type ordering scheme. The mechanism for the formation of the $P$2 $G$-type state in HMO was attributed to an intersite charge transfer between $A'$ and $B$ sites, caused by $1/3$ of the $A'$ sites undergoing a change in formal valence from Mn$^{3+}$ to Mn$^{2+}$ to result in a 1:1 ratio of Mn$^{3+}$:Mn$^{4+}$ residing on the $B$ sites. A key consequence of the intersite charge transfer and the formation of the $P$2 state is that ferroelectricity should be possible based on symmetry arguments, giving rise to a polar ground state. While polar ground states, albeit still uncommon, have been confirmed so far in other manganite quadruple perovskites such as BiMn$_7$O$_{12}$,\cite{belik2017complex, khalyavin2020emergent} the $G$-type charge and orbital order state, as far as we are aware, has not been observed before in any other manganite perovskites.   

Motivated by the richness in orbital ordering states in these quadruple perovskites, we probe the evolution of charge and orbital order phenomena in the hole-doped quadruple perovskite Hg$_{1-x}$Na$_x$Mn$_3$Mn$_4$O$_{12}$ (HNMO). Through the use of careful refinements of structural models against high-resolution synchrotron powder x-ray diffraction (PXRD) data, utilising symmetry-motivated approaches, we show that the novel polar $G$-type charge and orbital ordered state is robust to formation across the range of compositions studied here. Through invariants analysis we attribute the formation of this novel state to the coupling of distortions describing structural and charge and orbital degrees of freedom with that of ferroelectric polarization. Finally, we observe for optimal hole doping conditions $<$Mn$_B$$>$ $\approx$ $+$3.375 in HNMO, coincident with that giving maximal response in CMR in LCMO and LPCMO, an intrinsic electronic phase separation of $G$-type and OO:CD-type states occurs. This nature of phase separation in optimally doped HNMO further highlights the functional importance of electronic segregation that occurs in the manganite perovskites, proving vital to understand how this can be exploited in tuning the design and optimization of technologically relevant physical properties.  

\section{Experiment and data analysis}

Compositions of HNMO ($x$ $=$ 0.05, 0.1, 0.15, 0.2, 0.3, 0.4, 0.5 were synthesized by high-pressure high-temperature (HPHT) solid-state synthesis methods. High-resolution synchrotron powder x-ray diffraction (PXRD) techniques were used to characterize these compositions using different synchrotron sources for different types of data collection. In each case samples were loaded into 0.1 mm diameter quartz or borosilicate capillaries and data were collected in Debye-Scherrer geometry. Variable-temperature PXRD data for each composition were collected on the high-resolution PXRD beamline 19A at the Taiwan Photon Source (TPS) over a temperature range 100 K $\leq$ $T$ $\leq$ 700 K. Two different blocks of datasets were collected: `high' and `low' temperature. Different x-ray energies were used and their wavelengths were determined through Rietveld refinements of a LaB$_6$ NIST standard against recorded PXRD data, these energies and wavelengths were: 20 keV ($\lambda$ $=$ 0.6199145(4) Å) and 16 keV ($\lambda$ $=$ 0.7748902(4) Å). Higher resolution ($\delta$$d$/$d$) measurements were collected on the high-resolution PXRD beamline ID22 at the European Synchrotron Radiation Facility (ESRF) at temperatures $T$ $=$ 300 K and 10 K for each HNMO composition. The x-ray energy used was 35 keV which corresponded to a wavelength $\lambda$ $=$ 0.354293(9) Å through Rietveld refinements of a NIST Si standard against its recorded PXRD data. 

Rietveld refinements of models against HNMO PXRD data allowing the Hg occupancy to refine freely consistently reached values of $\sim$ 90$\%$ of the nominal occupancy, in keeping with what we have observed in our previous work for HMO.\cite{chen2018improper} However, the consistent and systematic variation of lattice degrees of freedom across the series, as discussed throughout the text, indicates the nominal Na doping has been successful. We had previously refined HMO in a $Pnn$2 space group since this produced the same quality of fit but with significantly less refinable parameters compared to a full $P$112 structural model. However, due to the superior resolution of data obtained on ID22 this has allowed us to refine the additional monoclinic lattice distortion associated with $P$2 symmetry. Nonetheless, for the internal degrees of freedom we choose to still use these $Pnn$2 pseudo-symmetry constraints, in keeping with our previous work, and we will refer to this model as the $P$2 model throughout the remainder of the text. Rietveld refinements of each different HNMO structural model were performed in the basis of symmetry-adapted distortions that transform as the relevant active irreps in each phase. All irreps are given with respect to the $AB$O$_3$ $Pm\bar{3}m$ perovskite structure with the $A$ site at the origin and a cubic lattice parameter $a$ $=$ 3.72412 Å through which symmetry-adapted strains are calculated in reference to. These choices facilitate the widest general comparison of results to other different perovskite systems, and are consistent with our previous works. The full list of active irreps in each structural model, along with their order parameter directions (OPDs), are given in the Supplementary Information (Table S1). These irreps were obtained through the use of the ISOTROPY package ISODISTORT.\cite{Campbell1} For sake of conciseness, in the main text only strain irreps contain their OPDs while irreps due to atomic displacements are given without their OPD. Irreps are given with the labels $|Q(K_n^{+/-})|$ where $Q$ denotes the amplitude of the irrep, given as aristotype cell-normalized ($A_p$) values in units of Å, and $K_n^{+/-}$ denotes the label of the irrep itself. For a more detailed discussion on irreps and their associated OPDs the reader is directed to the work of Senn and Bristowe.\cite{senn2018group}

\section{Results and discussion}

\subsection{\label{sec:level2}Probing Phase Transition Temperatures}

\begin{figure}[!t]
\includegraphics{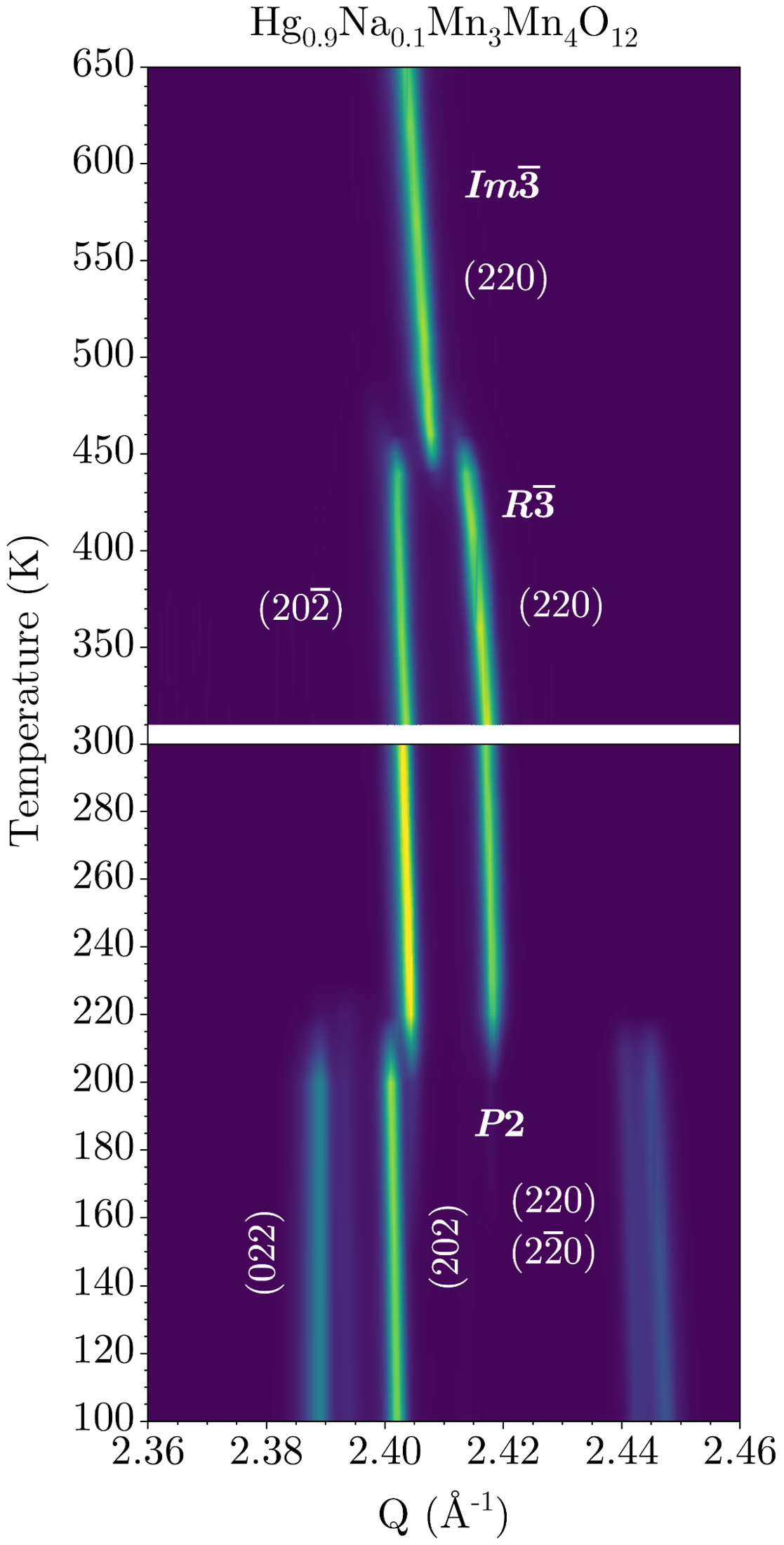}
\caption{\label{heatmaps} Variable-temperature synchrotron PXRD data, obtained on 19A TPS, demonstrated as a linearly-interpolated heatmap for the HNMO composition $x$ $=$ 0.1 over the temperature range $\approx$ 100 K - 650 K. White regions in the heatmap indicate temperature regions where data were not collected between `high' and `low' temperature datasets. The cubic-trigonal-monoclinic phase transitions are apparent $via$ splitting of the $Im\bar{3}$ (220) reflection. The $Im\bar{3}$ phase contains a cubic lattice parameter $a$ $\approx$ 7.4 Å and is double that of the $Pm\bar{3}m$ aristotype cubic lattice parameter from which symmetry-adapted strains are derived from.}
\end{figure}

Variable-temperature PXRD data obtained using the high-resolution PXRD beamline 19A at TPS for the HNMO compositions $x$ $=$ 0.05, 0.1, 0.2, 0.3, 0.4, 0.5 between 100 K $\leq$ $T$ $\leq$ 700 K are shown in Figure S1, with a representative case of $x$ $=$ 0.1 given in Figure \ref{heatmaps}. This $Q$ range highlights the evolution of the cubic (220) reflection splitting, indicating the presence of the $Im\bar{3}$ to $R\bar{3}$ and $R\bar{3}$ to $P$2 phase transitions. The series of phase transitions results in corresponding changes of lattice parameters given in Table S2 - S7. The cubic-to-trigonal transition temperature decreases as a function of increased hole doping from $T$ $\approx$ 460 K for $x$ $=$ 0.05 to $T$ $\approx$ 310 K for $x$ $=$ 0.4, where for $x$ $=$ 0.5 this respective transition does not occur. The trigonal-to-monoclinic phase transition, on the other hand, appears to exhibit a more subtle hole doping dependence on the transition temperature which decreases from $T$ $\approx$ 220 K for $x$ $=$ 0.05 to $T$ $\approx$ 180 K for $x$ $=$ 0.5. For the compositions with $x$ $=$ 0.4 and 0.5, however, a coexisting phase forms with the same $C$2/$m$ crystal structure as observed in NaMn$_3$Mn$_4$O$_{12}$.\cite{streltsov2014jahn, chen2021striping} The nature of this phase coexistence will be further discussed in Section \ref{10K-refinements}.

\begin{figure}[!t]
\includegraphics{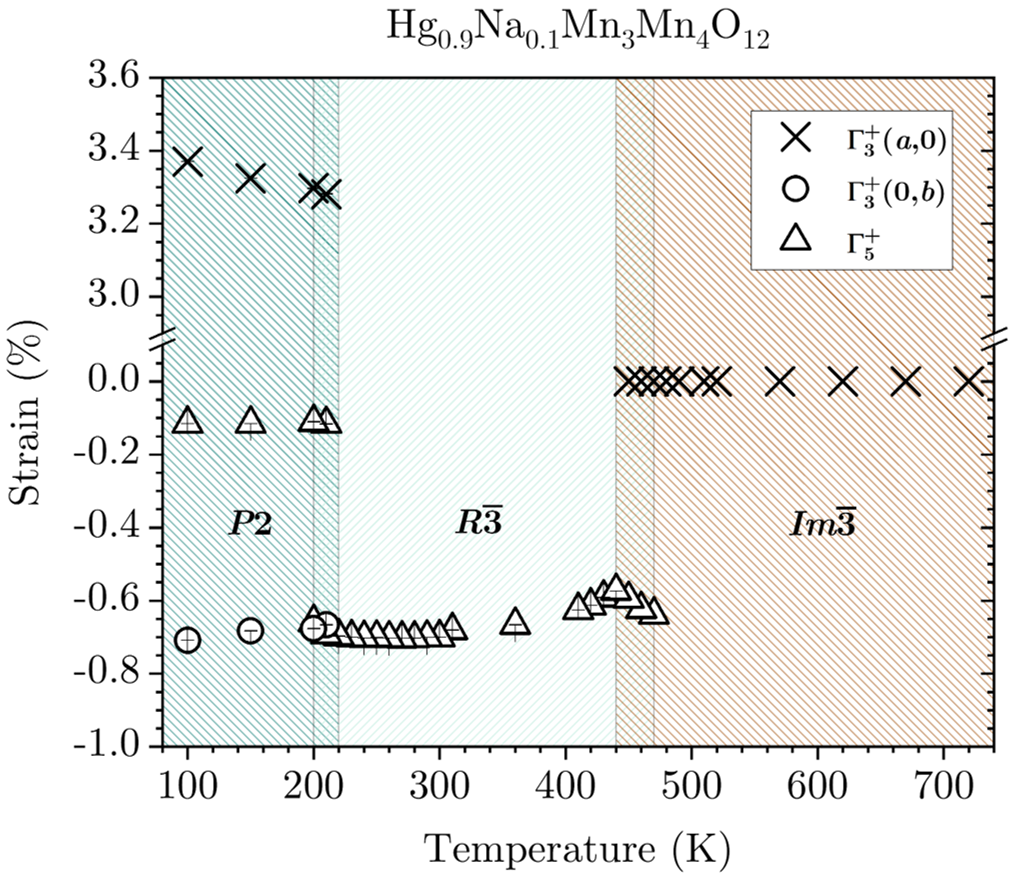}
\caption{\label{VT-strain} Variation of tetragonal, orthorhombic and shear strain ($\Gamma_3^+$($a$,0), $\Gamma_3^+$(0,$b$) and $\Gamma_5^+$ respectively) of the HNMO composition $x$ $=$ 0.1 obtained from Rietveld refinements against structural models against data shown in Figure \ref{heatmaps}. Regions illustrating the presence of $Im\bar{3}$, $R\bar{3}$ and $P$2 phases, along with phase-coexistence regions, are given. Note the discontinuity of the tetragonal strain $\Gamma_3^+$($a$,0) contains values of over an order of magnitude greater compared to that of the orthorhombic strain $\Gamma_3^+$(0,$b$). Since the strain irrep $\Gamma_5^+$ contains a different set of OPDs depending on whether it is due to the $R\bar{3}$ or $P$2 phase, only the general irrep label is given. Errors are given for each data point, and in some cases they are smaller than the size of the point itself.}
\end{figure}

Evolution of the tetragonal, orthorhombic and shear lattice strains ($\Gamma_3^+$($a$,0), $\Gamma_3^+$(0,$b$) and $\Gamma_5^+$ respectively) for the HNMO composition $x$ $=$ 0.1 is shown in Figure \ref{VT-strain}. All other compositions are given in Figure S2. Each of these three strains are necessarily zero for the $Im\bar{3}$ phase, and for the $R\bar{3}$ phase only $\Gamma_5^+$($a$,$a$,$a$) is active. In both $P$2 and $C$2/$m$ phases each of the symmetry-breaking strains $\Gamma_3^+$($a$,0), $\Gamma_3^+$(0,$b$) and $\Gamma_5^+$($a$,0,0) are active. Discontinuities in each of the symmetry-adapted strains are suggestive of a first-order nature to both phase transitions, coupled with regions of phase coexistence. Despite the hole doping dependence of the phase transition temperatures, the range over which phase coexistence occurs between the `disordered $C$-type' state ($R\bar{3}$) and polar $G$-type state ($P$2) appears to be robust across the HNMO series. Perhaps most noteworthy, the tetragonal strain in the $P$2 phase for each composition is consistently and significantly greater than both orthorhombic and shear strains ($i.e.,$ as shown in Figure \ref{VT-strain} at 100 K: $\Gamma_3^+$($a$,0) $\approx$ 3.4$\%$, $\Gamma_3^+$(0,$b$) $\approx$ $-$0.7$\%$, $\Gamma_5^+$($a$,0,0) $\approx$ $-$0.1 to 0$\%$). This tetragonal elongation of lattice strain, and its directionality, arises due to the inherent coupling to the JT-long axis of rock-salt charge-ordered Mn$^{3+}$ on the $B$ sites which all point along the $c$-axis. A more complete analysis of the internal degrees of freedom associated with the two different low temperature monoclinic phases and the room temperature trigonal/cubic state for each composition benefit from data obtained on ID22 at the ESRF.  

\subsection{\label{300K-refinements} Tuning ambient temperature trigonal and cubic states}

\begin{figure}[!b]
\includegraphics{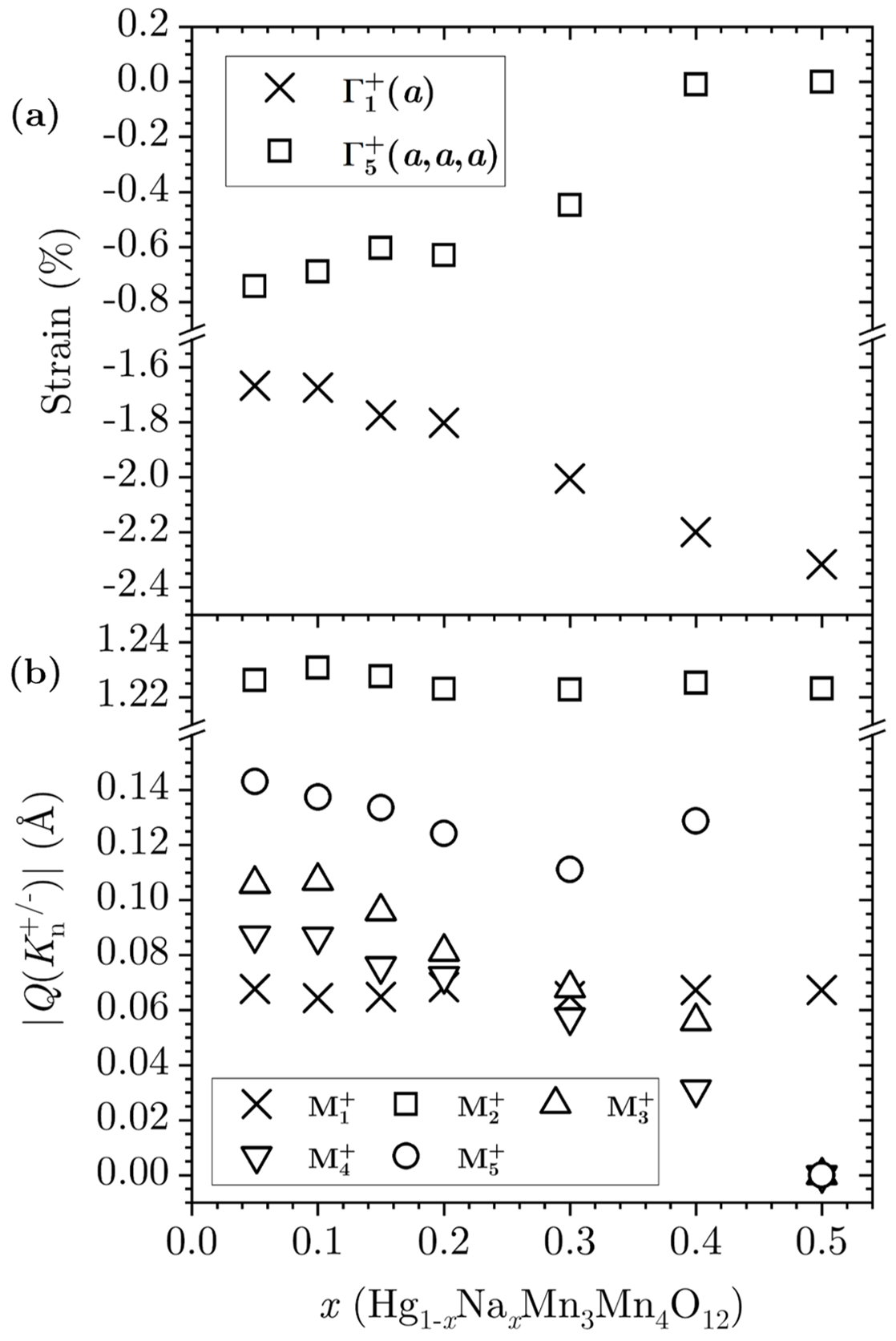}
\caption{\label{RT-modes} Extracted strain (a) and distortion modes (b) from Rietveld refinements of structural models, $R\bar{3}$ or $Im\bar{3}$ phases, against ID22 ESRF data at $T$ $=$ 300 K for each studied HNMO composition. Each M-point mode is active for the $R\bar{3}$ phase (compositions $x$ $=$ 0.05, 0.1, 0.15, 0.2, 0.3, 0.4), whereas only M$_1^+$ and M$_2^+$ are active in the $Im\bar{3}$ phase (composition $x$ $=$ 0.5). Errors are omitted here, but can be found in an identical figure in Figure S4.}
\end{figure}

Lattice strain and atomic distortion modes extracted from Rietveld refinements of $R\bar{3}$ structural models ($Im\bar{3}$ for $x$ = 0.5) against data obtained on ID22 at 300 K are shown in Figure \ref{RT-modes}. Rietveld refinement profiles for each HNMO composition are given in Figure S3, which indicate that a single phase is sufficient to provide a good quality fit to the data (with the exception of $x$ $=$ 0.5). The symmetric volume strain, $\Gamma_1^+$($a$), gives an expected linear increase in its magnitude (increasingly negative consistent with a decreasing unit cell volume) with increased hole doping due to the formation of greater proportions of the smaller ionic radii Mn$^{4+}$ from larger ionic radii high spin Mn$^{3+}$ (0.53 Å $vs$ 0.645 Å respectively in an octahedral coordination environment).\cite{shannon1976revised} This linearity is therefore indicative that we are sufficiently close to the nominal Na doping level. Shear strain, $\Gamma_5^+$($a$,0,0), demonstrates a systematic decrease in its magnitude towards zero strain for increased hole doping towards $x$ $=$ 0.4. The decrease of this shear strain indicates that HNMO compositions undergo a continuous transition from trigonal to cubic states. From analysis of the internal degrees of freedom, one finds that all of these distortion modes transform as irreps associated with the M-point of the simple $AB$O$_3$ perovskite structure. Distortions describing $A$ site cation order and the $a^+$$a^+$$a^+$ $B$O$_6$ octahedral rotation mode (from Glazer notation\cite{glazer1972classification}), M$_1^+$ and M$_2^+$ respectively, are essentially constant with hole doping, demonstrating the quadruple perovskites' ability to `lock in' a single magnitude of octahedral rotations. Distortions describing the degree of `disordered $C$-type' orbital order, M$_3^+$, $B$ site cation (charge) order, M$_4^+$, and a shearing of $B$O$_6$ octahedra, M$_5^+$, all demonstrate systematic decreases to zero magnitude for increased hole doping, as these modes are all necessarily zero by symmetry in the $Im\bar{3}$ phase. Hence, the internal degrees of freedom provide a further indication of the continuous nature of the change of phases from trigonal to cubic symmetries as a function of hole doping.

An additional, and perhaps most crucial, feature of these data is that the extracted microstrain from modelling the peak profiles in the Rietveld refinements occur within the range $\sim$ 0.01$\%$ to 0.036 $\%$ (Figure S5). The reported microstrain and macrostrain, the latter derived from symmetry-adapted strains, may be used to distinguish whether any phase separation due to subsequent phase transitions upon cooling is caused by electronic or chemical effects. For example, given the microstrain for any HNMO composition is significantly smaller than the difference of macrostrain between two adjacent HNMO compositions, this places stringent criteria on the chemical homogeneity of the sample, implying any phase separation induced on cooling is intrinsically electronic in character. The significance of this point will become apparent in Section \ref{10K-refinements}. 

\subsection{\label{10K-refinements} Polar $G$-type ground states and electronic phase separation}

\begin{figure}[!t]
\includegraphics{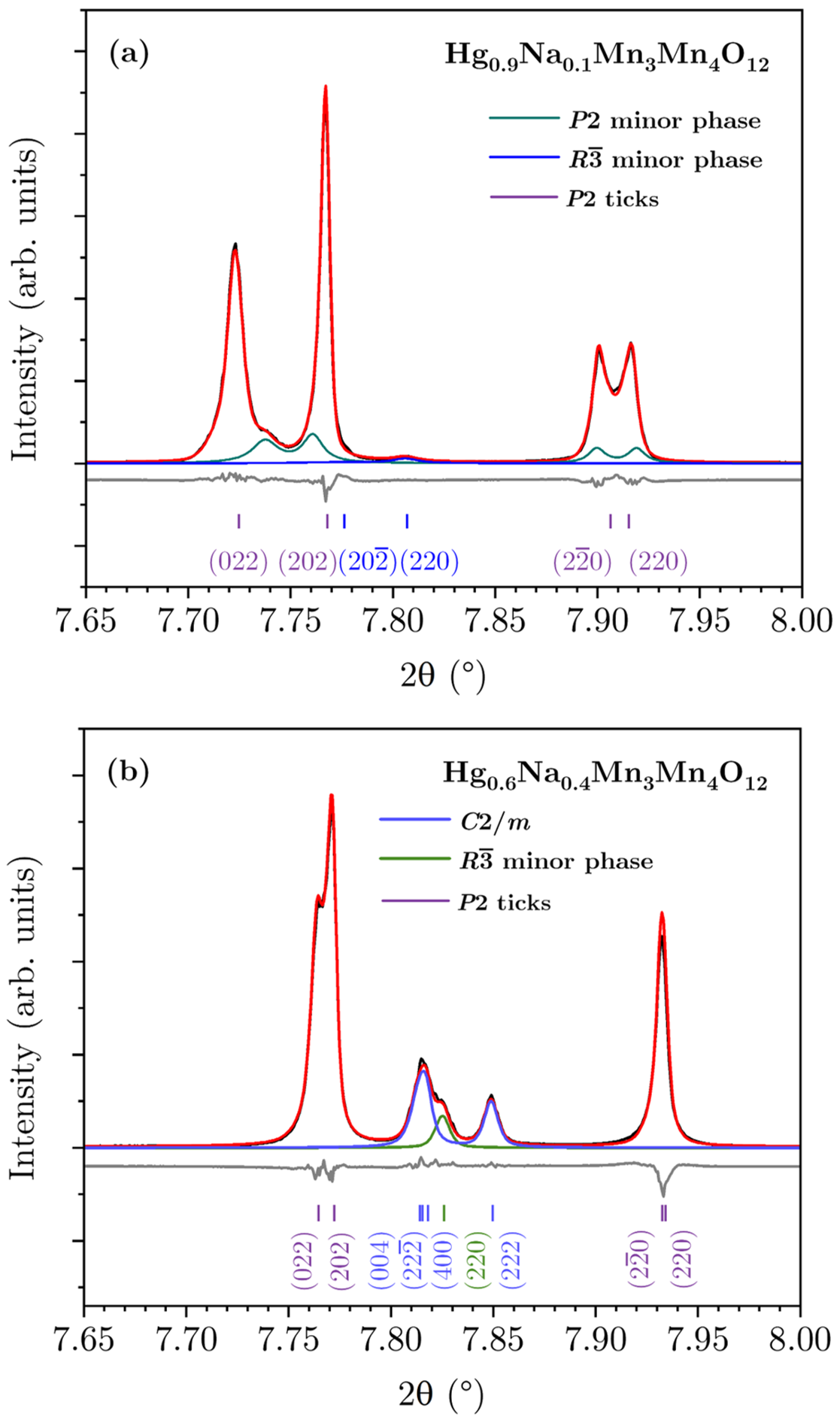}
\caption{\label{Na010-Na040-10K} Profiles of Rietveld refinements of structural models against ID22 ESRF data at $T$ $=$ 10 K for the HNMO compositions (a) $x$ $=$ 0.1 and (b) $x$ $=$ 0.4. Black, red and grey traces correspond to observed, calculated and difference profiles respectively. For (a), the component of the $R\bar{3}$ phase contains a phase fraction of $\approx$ 3$\%$. For (b), the full phase fractions each phase constitute are: $P$2 $=$ 76$\%$, $C$2/$m$ $=$ 18$\%$ and $R\bar{3}$ $\approx$ 4$\%$. The remaining fraction corresponds to trace amounts of Mn$_2$O$_3$ (not observable in this 2$\theta$ range).}
\end{figure}

Profiles of Rietveld refinements of structural models against ID22 ESRF data at $T$ $=$ 10 K are shown in Figure S6. A zoomed-in region of the profile for the HNMO composition $x$ $=$ 0.1, shown in Figure \ref{Na010-Na040-10K}(a), highlights the monoclinic splitting of the cubic (220) reflection into (2$\bar{2}$0) and (220) reflections demonstrating the formation of the $P$2, polar $G$-type charge and orbital ordered state as observed in HMO. This type of monoclinic splitting occurs for each HNMO composition studied here. A second $P$2 phase is fit to the data in order to improve the quality of the fit, containing a reduced tetragonal strain state, which is likely associated with residual strained regions at domain boundaries. A further small phase fraction ($\approx$ 3$\%$ from Rietveld refinements) of an $R\bar{3}$ phase is also fit to the data which is the result of small, strained, domains of the $R\bar{3}$ phase which have been unable to relax during the nucleation of the $P$2 phase. Both of these features are generally observed for all HNMO compositions studied here, but this does not affect the extraction of accurate structural degrees of freedom from our refinements. 

Strikingly for the HNMO composition $x$ $=$ 0.4, shown in Figure \ref{Na010-Na040-10K}(b), corresponding to the optimal hole doping $<$Mn$_B$$>$ $\approx$ +3.375 as observed in LCMO/LPCMO giving maximal CMR responses, the $P$2 state still occurs but with the formation of a second coexisting phase of $C$2/$m$ symmetry known to exist for the ground state structures of NaMn$_3$Mn$_4$O$_{12}$ and also intermediate hole doped Ca$_{1-x}$Na$_x$Mn$_3$Mn$_4$O$_{12}$.\cite{chen2021striping} The same phenomenon is also observed for the composition $x$ $=$ 0.5 (Figure S6). The relative phase fractions obtained from Rietveld refinements between $P$2:$C$2/$m$ phases in both of these compositions are determined as: $x$ $=$ 0.4, $\sim$ 80$\%$:20$\%$ and $x$ $=$ 0.5, $\sim$ 20$\%$:80$\%$. We attribute the mechanism for phase coexistence of these HNMO compositions to electronic, rather than chemical, phase separation for a number of crucial reasons. Data obtained at $T$ $=$ 300 K (Figure S3), which is above the ordering temperature of the ground states, produced a good quality fit using a single structural model for all compositions adopting the $R\bar{3}$ state. The exception to this is the $x$ $=$ 0.5 HNMO composition adopting the $Im\bar3$ state where a second equivalent phase with similar lattice parameters was required to provide a satisfactory fit. If chemical phase separation were to occur then multiple unique phases with distinctly different lattice strains would be resolvable, which we do not observe in our data. Furthermore, the exceptionally low microstrain of the samples at $T$ $=$ 300 K, before the observed electronic phase coexistence, precludes any significant chemical phase separation. Microstrains of our samples are below 0.036 $\%$ as a maximum obtained value (Figure S5). The chemical heterogeneity implied by the distribution of macrostrains relatable to this microstrain is equivalent to a change in nominal Na doping of $\delta$$x$ $\sim$ 0.01, $i.e., $ 1/10$^{th}$ of the variation in the symmetric volume strain $\Gamma_1^+$($a$) between neighboring compositions such as $x$ $=$ 0.3 and 0.4. We may thus conclude the observed phase separation is of electronic and not chemical origin. These intrinsic electronic phase separation phenomena are known to occur in many systems exhibiting metal-to-insulator transitions such as in the superconducting cuprates and in the CMR manganite perovskites,\cite{tidey2022pronounced, uehara1999percolative} further highlighting the importance of its effect in the generation of exotic physical properties. 

\begin{figure}[!t]
\includegraphics{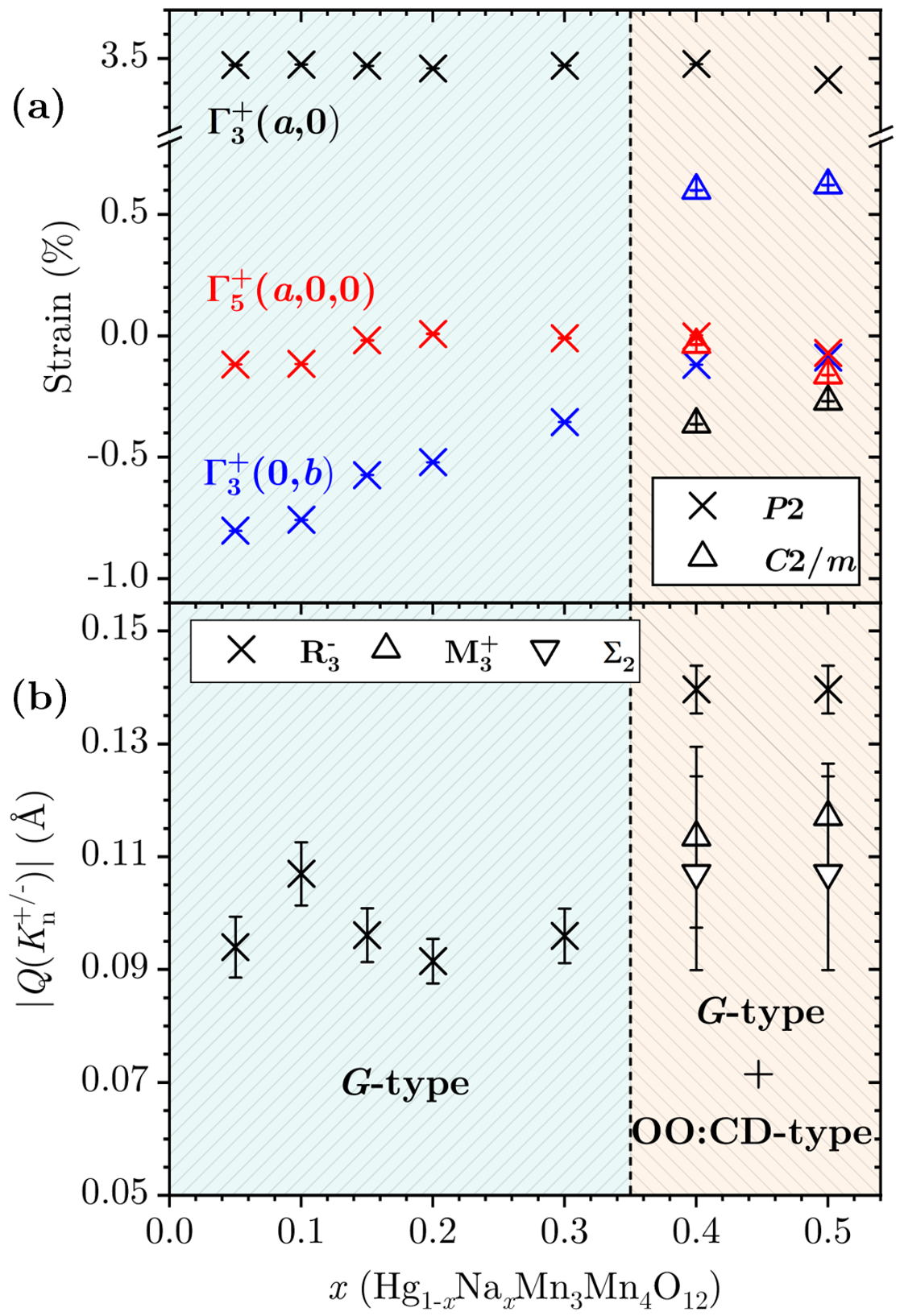}
\caption{\label{ground-state-strain-modes} (a) Variation of tetragonal, orthorhombic and shear strain ($\Gamma_3^+$($a$,0), $\Gamma_3^+$(0,$b$) and $\Gamma_5^+$($a$,0,0) respectively) of each HNMO composition for both $P$2 and $C$2/$m$ phases. (b) Variation of structural distortions that transform as irreps describing the emergence of different charge and orbital ordered state of each HNMO composition. In each panel strain and distortion modes were extracted from Rietveld refinements of structural models against ID22 ESRF data at $T$ $=$ 10 K. Regions illustrating windows in which $G$-type change and orbital order, and phase coexistence between $G$-type and OO:CD-type states exist are shown by blue and orange shading, respectively. Errors are given for each data point, and in some cases they are smaller than the size of the point itself.}
\end{figure}

With the nature of the ground state phase coexistence now established for optimally doped HNMO compositions, we look to investigate their structural variation with respect to both lattice strain and internal degrees of freedom. Extracted tetragonal, orthorhombic and shear strain of both relevant $P$2 and $C$2/$m$ phases from these refinements are given in Figure \ref{ground-state-strain-modes}(a), and key structural distortions transforming as irreps that result in the different charge and orbital ordered states in Figure \ref{quadruple-perovskite-ordering-schemes} (R$_3^-$, M$_3^+$ and $\Sigma_2$) are shown in Figure \ref{ground-state-strain-modes}(b). All other distortion modes are given in Figure S8 and Table S8. 

Firstly, we will focus on the structural evolution of the polar $G$-type $P$2 state. The variation of shear strain and orthorhombic strain towards zero as a function of increased hole doping indicates the reduction of monoclinicity and orthorhombicity, respectively, across the HNMO series. This is most apparently observed by a reduction of the splitting of (2$\bar{2}$0) and (220) reflections and a reduction of the splitting of (022) and (202) reflections, respectively, shown in Figure \ref{Na010-Na040-10K}. Considering the distortion mode evolution attributing to charge and orbital order structural distortions, only the irrep R$_3^-$ is active for the HNMO composition range 0.05 $\leq$ $x$ $\leq$ 0.3, supporting the presence of $G$-type charge and orbital order. This character of $G$-type orbital order has been previously observed in a number of vanadate oxide perovksites\cite{sage2006evidence, ritter2016crystallographic, saha2017coexistence} as well as $A$CrF$_3$ and $A$CuF$_3$ ($A$ $=$ cation) perovskites.\cite{margadonna2006cooperative, zhou2011jahn} The nature of the $P$2 phase here demonstrates the robust formation of the $G$-type ordered state across the hole doping regime $<$Mn$_B$$>$ $=$ +3.25 to +3.325 in HNMO quadruple perovskites. Furthermore, because the $P$2 space group is polar, by symmetry arguments these compositions adopting the $G$-type state should have potential ferroelectric activity. We have previously identified the emergence of ferroelectric polarization in the $P$2 $G$-type state through an $A'$-$B$ intersite charge transfer which acts to form a 1:1 ratio of Mn$^{3+}$:Mn$^{4+}$ on the $B$ site and produce a 1:2 Mn$^{2+}$:Mn$^{3+}$ ratio of the $A'$ site formal valences.\cite{chen2018improper} Similar charge transfer phenomena have been observed in other quadruple perovskites such as LaCu$_3$Fe$_4$O$_{12}$,\cite{long2009temperature} albeit it is usually the case that the $A'$ site is robust to its formal valence.

Corroborating these changes in formal valences, bond valence sum (BVS) analysis of the Hg $A$, Mn $A'$ and Mn $B$ sites was performed for each HNMO composition. Full data are given in Figure S9. The Hg $A$ sites give consistent BVS values of $\sim$ +2, the Mn $B$ sites produce a 1:1 splitting of +3.2:+3.6 (formally given as +3:+4) and the Mn $A'$ sites produce a range of BVS values from +2.85 to +3.1. It should be noted that BVS analysis techniques tend to underestimate the degree of charge separation in charge-ordered systems. This observation has been reported for various perovskites undergoing charge ordering such as Pr$_{0.5}$Ca$_{0.5}$MnO$_3$, YNiO$_3$ and TbBaMn$_2$O$_6$.\cite{goff2004charge, alonso1999charge, williams2002alternative} Hence, using exact values should not be entirely conclusive to determine the exact formal valence of a particular site. Instead, however, looking at the ratio between `low' and `high' BVS values for Mn cations residing on either $A'$ or $B$ sites provides a more clear indication of charge transfer processes that may occur. At higher hole doping (for $x$ $=$ 0.3 and above) the predicted 1:2 lower:higher BVS ratio of the Mn $A'$ sites is observed, and we take one of these compositions ($x$ $=$ 0.3) for further structural analysis. Within this structure we observe a short contact Mn-O bond distance of $\approx$ 2.4 Å that occurs for 1/3 of the square planar $A'$ sites consistent with our previous work on HMO.\cite{chen2018improper} This short contact bond at this bond length value has been further observed in other quadruple perovskites containing Mn$^{2+}$ $A'$ sites.\cite{tohyama2013valence, aimi2014high} 

\begin{figure}[!t]
\includegraphics{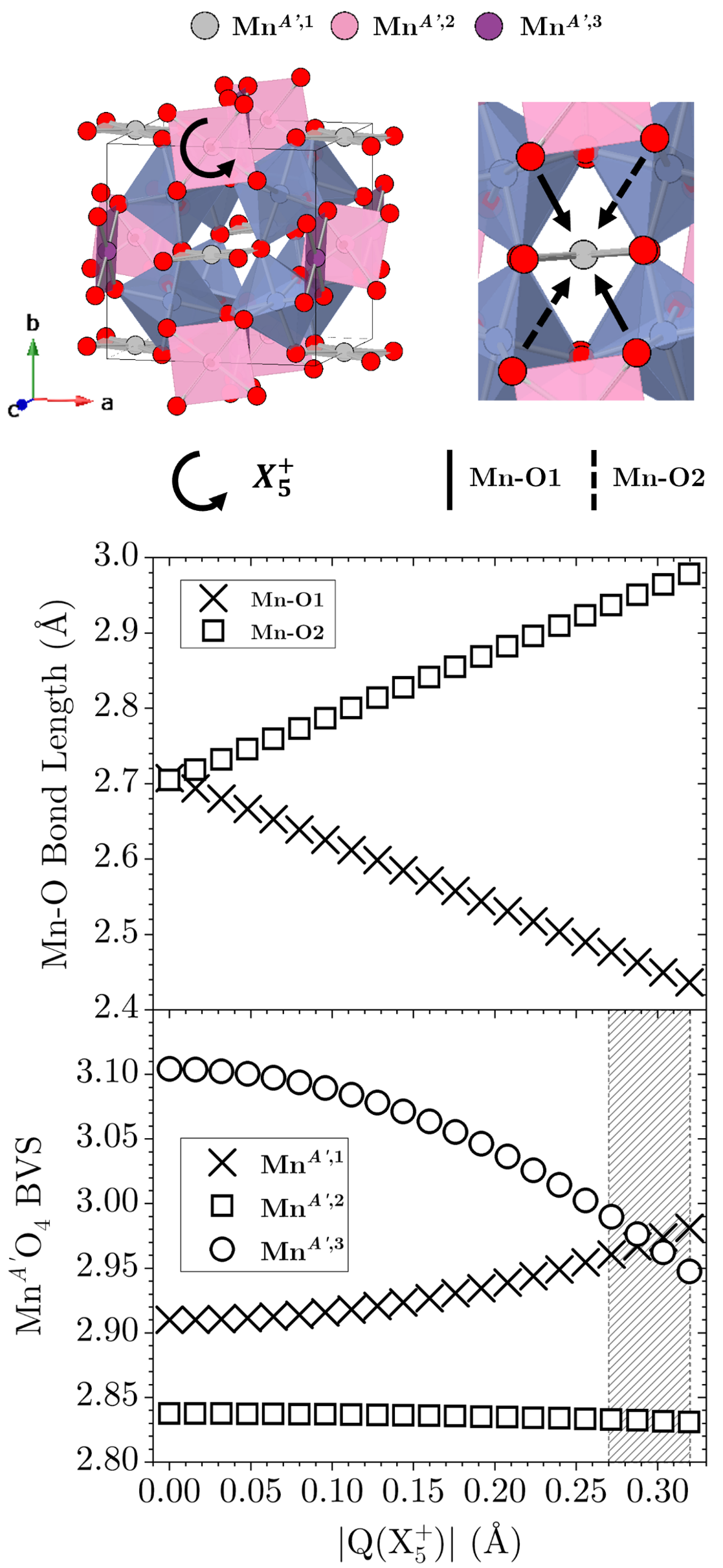}
\caption{\label{X5p_rotation} The effect of tuning the irrep X$_5^+$ on the crystal structure of the polar $P$2 $G$-type phase $via$ rigid $A$'O$_4$ square planar rotations about the $c$-axis. $A$ sites are omitted for clarity. The maximum amplitude of X$_5^+$ is the value that is adopted for the HNMO composition $x$ $=$ 0.3, and this value is systematically reduced by 5$\%$ increments. The shaded region indicates the optimal range of X$_5^+$ distortion mode amplitude values required to satisfy a 1:2 Mn$^{2+}$:Mn$^{3+}$ formal valence ratio necessary for $A'$ site charge transfer to occur in HNMO.}
\end{figure}

From the perspective of symmetry-motivated analysis, this formation of the $\approx$ 2.4 Å Mn-O bond in 1 out of the 3 distinct $A'$ sites arises due the structural distortion transforming as X$_5^+$ which causes a rigid rotation of each $A'$O$_4$ square planar unit about the $c$-axis. This is equivalent to the irrep H$_4^+$ with respect to the $Im\bar{3}$ perovskite aristotype structure, which we had identified before.\cite{chen2018improper} We illustrate in Figure \ref{X5p_rotation} the structural significance of X$_5^+$ by taking the refined model of the HNMO composition $x$ $=$ 0.3 (X$_5^+$ $\approx$ 0.32 Å) and systematically reduce this to zero in small increments whilst keeping all other distortion modes fixed. We then plot in Figure \ref{X5p_rotation} how the bond distance and different $A'$ site BVS values vary as a function of X$_5^+$. While the Mn-O bond distances within the plane of the $A'$O$_4$ square planar coordination environment are essentially invariant to the action of X$_5^+$, the additional short and long Mn-O bond distances orthogonal to the plane show an apparent evolution with decreasing mode amplitude (Figure \ref{X5p_rotation}). Additionally, we observe that `low' and `high' $A'$ site BVS values maintains a 1:2 ratio for a band of X$_5^+$ amplitudes close to the globally refined value, which then deviates as X$_5^+$ decreases further towards zero. Both this rotation behavior and the distortion transforming as X$_5^+$ are not active in any of the $Im\bar{3}$, $R\bar{3}$ or $C$2/$m$ phases studied here and in other quadruple perovskites. Hence, we indicate that it is this distortion that gives rise to the $A'$ site charge order process occurring in HNMO, attributable to the formation of ground state polarization in the $P$2 $G$-type charge and orbital ordered state. By expanding the Landau free energy of the $P$2 state we find through invariants analysis\cite{Hatch1} the general 4$^{th}$ order linear term R$_3^-$X$_5^+$M$_2^+$$\Gamma_4^-$ emerges. Here, since M$_2^+$ is always active due to the cation ordering and associated octahedral rotations in these quadruple perovskites, we may view this as a trilinear coupling mechanism between order parameters that couple $B$ site charge and orbital order (R$_3^-$) with $A$ site charge order (X$_5^+$) to an improper ferroelectric polarization ($\Gamma_4^-$). 

This mechanism therefore indicates design criteria which allows ferroelectricity to emerge through the coupling of structural, charge and orbital degrees of freedom. Crucially, if one could grow these materials under epitaxial compressive strain, not only would this promote orbital ordering $via$ coupling to macroscopic tetragonal strain, it would also produce an out-of-phase polarization that is optimal for any device-based application. Clearly, the design of such devices which would harness strong magnetoelectric coupling due to intrinsic coupling between polarization and orbital order in these materials would be very favorable. To this end, the determination of the magnetic structure of the $P$2 phase and how it intimately ties in to the rich array of unique charge and orbital order textures highlighted here will be of great interest for future investigation.        

We now turn our attention to the $C$2/$m$ phase that appears in the phase-coexisting HNMO compositions $x$ $=$ 0.4 and 0.5. The relevant structural distortion modes describing $C$-type and $CE$-type charge and orbital order states (M$_3^+$ and $\Sigma_2$ respectively) emerge and their refined values are equal to each other within error (Figure \ref{ground-state-strain-modes}). This equal ratio of M$_3^+$:$\Sigma_2$ distortion mode amplitudes is a key primary signature for the presence of the novel OO:CD-type state we previously discovered in Ca$_{1-x}$Na$_x$Mn$_3$Mn$_4$O$_{12}$ perovskites around the same optimal doping level.\cite{chen2021striping} The coherent superposition of these distortion modes necessarily discriminates it from either $C$-type or $CE$-type states existing in this $C$2/$m$ phase. A further structural ingredient illustrating OO:CD-type state formation is the presence of a pseudo-tetragonal strain state, shown in Figure \ref{ground-state-strain-modes}, and its invariance across this optimal doping range.\cite{chen2021striping} While both this state and the previously identified $P$2 $G$-type state contain a pseudo-tetragonal lattice distortion, these are of notably different magnitudes (the $G$-type state being much greater than the OO:CD-type state) and so they are not directly related to each other. This then further indicates that the phase coexistence is one of electronic phase separation rather than chemical separation, giving a physical basis for the strongly first order transition and hence the observed electronic phase separation. Ultimately, the OO:CD-type state occurs in HNMO compositions with $<$Mn$_B$$>$ $\approx$ +3.375, whilst coexisting with the polar $P$2 $G$-type state. While the coexistence of charge and orbital ordered $and$ disordered metallic states through electronic phase separation in perovskites is a reasonably common observation, the coexistence of phases exhibiting distinct orbital ordering modes is evidently more rare. Notable examples where this type of coexistence occurs are within the vanadate perovskites $A$VO$_3$ exhibiting coexisting $C$-type (M$_3^+$) and $G$-type (R$_3^-$) states.\cite{sage2006evidence, ritter2016crystallographic, saha2017coexistence} In the case of HNMO, however, three distinct orbital ordering modes coexist within the electronic phase separated region: two coexist and interfere coherently (M$_3^+$ and $\Sigma_2$) to give rise to a OO:CD-type state, and the other (R$_3^-$) exists alone giving rise to a spatially-separated $G$-type state. 

\begin{figure}[!t]
\includegraphics{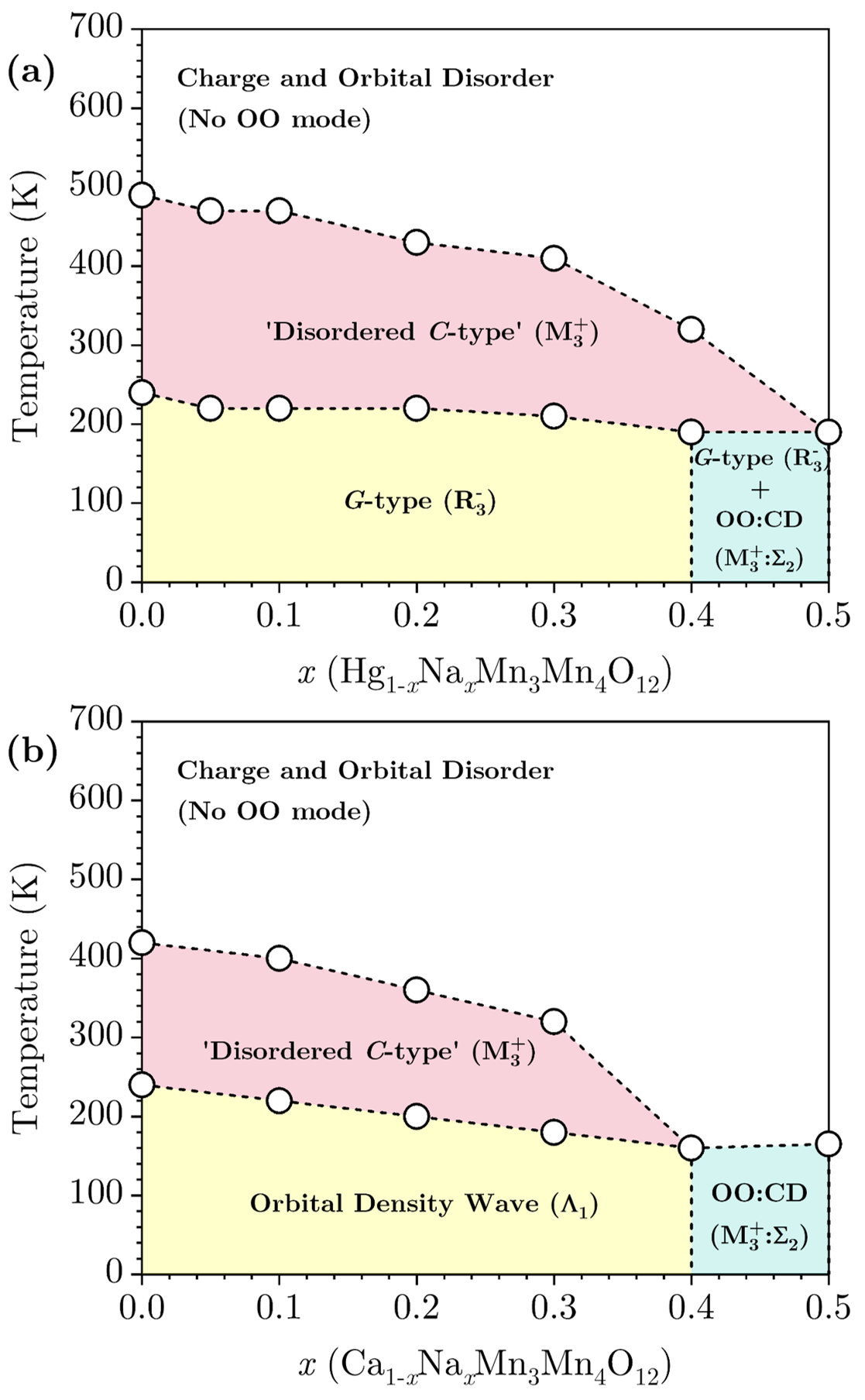}
\caption{\label{HNMO-phase-diagram} Charge and orbital order phase diagram of (a) HNMO compositions studied here and (b) Ca$_{1-x}$Na$_x$Mn$_3$Mn$_4$O$_{12}$ compositions we have previously studied.\cite{chen2021striping} Data points for the HNMO end member HgMn$_3$Mn$_4$O$_{12}$ are taken from our previous work.\cite{chen2018improper} Schematics for the different types of charge and orbital order are given in Figure \ref{quadruple-perovskite-ordering-schemes}. Irreps that transform as structural distortions responsible for the onset of the different ordered states are given in brackets. OO $=$ orbital order, CD $=$ charge disorder.}
\end{figure}

The culmination of results presented here produce the charge and orbital order phase diagram (Figure \ref{HNMO-phase-diagram}) of HNMO compositions studied here, which are compared directly to the corresponding hole doped Ca$_{1-x}$Na$_x$Mn$_3$Mn$_4$O$_{12}$ series previously studied.\cite{chen2021striping} Both of these systems illustrate how the doped quadruple perovskites facilitate a richer array of different charge and orbital order textures compared to the doped $AB$O$_3$ equivalents, forming an appealing system for further study in how other ordered configurations can be tuned and stabilized. Furthermore, these phase diagrams indicate that even the chemistries of Ca$^{2+}$ and Hg$^{2+}$, with near identical ionic radii, themselves cause substantial changes to the ground states of manganite quadruple perovskites, notably tuning the orbital density wave character of CaMn$_3$Mn$_4$O$_{12}$ to the $G$-type polar state of HgMn$_3$Mn$_4$O$_{12}$. How these states are tuned in the series of compositions spanning the solid solution Hg$_{1-x}$Ca$_x$Mn$_3$Mn$_4$O$_{12}$ will form a substantial body of future work.    

\section{Conclusion}

We present here the successful synthesis of the series of hole doped quadruple manganite perovskites Hg$_{1-x}$Na$_x$Mn$_3$Mn$_4$O$_{12}$ (HNMO) for members up to $x$ $=$ 0.5. We demonstrate the phase transition temperature-dependencies for the cubic-to-trigonal and trigonal-to-monoclinic transitions for each composition in the series, unique to quadruple perovskites containing Hg$^{2+}$ on the $A$ site. We obtain accurate structural models for these compositions at ambient temperature which show a compositionally-dependent trigonal-to-cubic phase transition, where the majority of the phases are fit with a single structural model. Regarding the ground state structures, we highlight the robust formation of the polar, $G$-type charge and orbital ordered state associated with HgMn$_3$Mn$_4$O$_{12}$ as a sole phase for compositions up to $x$ $=$ 0.3. Further hole doping results in the coexistence of two unique ordered states due to intrinsic electronic phase separation of $G$-type and OO:CD-type states, occurring at critical doping regimes consistent with that of LCMO and LPCMO manganite perovskites exhibiting a maximal CMR response. The intrinsic electronic nature of this phase separation has been validated through strain analysis of both microstrain and macrostrain quantities for compositions at temperatures both above and below the ground state charge and orbital order transition. In this instance we would speculate that other kinds of exotic physical properties might arise in these electronically phase separated HNMO compositions, akin to CMR observed in LCMO and LPCMO manganites, which remains a promising avenue of future study. 

Broadly, we indicate that it is possible to control the order parameters that underpin different unique charge and orbital ordered states, and how they might interfere coherently within domains and across domain walls on nanometer and micron-level length scales. One example strategy that might achieve this is through the use of exploiting strain fields to favor one particular ordering state/domain over the other. The insight obtained here is not just restricted to fundamental interest, but may have useful applications in materials design strategies for memristive technologies and neuromorphic computing.         

\begin{acknowledgments}

BRMT acknowledges the University of Warwick and the EPSRC for studentship and funding (EP/R513374/1), and thanks Catriona Crawford for assistance in the collection of ID22 ESRF synchrotron x-ray diffraction data and helpful discussions. MSS acknowledges the Royal Society for a fellowship (UF160265) and EPSRC grant "Novel Multiferroic Perovskites through Systematic Design" (EP/S027106/1) for funding. WTC acknowledges the National Science and Technology Council, Taiwan, for funding 111-2112-M-002-044-MY3, 112-2124-M-002-012, and Academia Sinica project number AS-iMATE-113-12, and the Featured Areas Research Center Program within the framework of the Higher Education Sprout Project by the Ministry of Education of Taiwan 113L9008. We acknowledge the Taiwan Photon Source (TPS) for synchrotron x-ray diffraction beamtime with proposal number 2023-1-166, the European Synchrotron Radiation Facilities (ESRF) for provision of synchrotron radiation facilities under proposal number HC-5203 (DOI: 10.15151/ESRF-ES-1210183160) and we would like to thank Ola Grendal and Andy Fitch for assistance and support using beamline ID22.

\end{acknowledgments}


\bibliography{references}

\begin{thebibliography}{41}%
\makeatletter
\providecommand \@ifxundefined [1]{%
 \@ifx{#1\undefined}
}%
\providecommand \@ifnum [1]{%
 \ifnum #1\expandafter \@firstoftwo
 \else \expandafter \@secondoftwo
 \fi
}%
\providecommand \@ifx [1]{%
 \ifx #1\expandafter \@firstoftwo
 \else \expandafter \@secondoftwo
 \fi
}%
\providecommand \natexlab [1]{#1}%
\providecommand \enquote  [1]{``#1''}%
\providecommand \bibnamefont  [1]{#1}%
\providecommand \bibfnamefont [1]{#1}%
\providecommand \citenamefont [1]{#1}%
\providecommand \href@noop [0]{\@secondoftwo}%
\providecommand \href [0]{\begingroup \@sanitize@url \@href}%
\providecommand \@href[1]{\@@startlink{#1}\@@href}%
\providecommand \@@href[1]{\endgroup#1\@@endlink}%
\providecommand \@sanitize@url [0]{\catcode `\\12\catcode `\$12\catcode
  `\&12\catcode `\#12\catcode `\^12\catcode `\_12\catcode `\%12\relax}%
\providecommand \@@startlink[1]{}%
\providecommand \@@endlink[0]{}%
\providecommand \url  [0]{\begingroup\@sanitize@url \@url }%
\providecommand \@url [1]{\endgroup\@href {#1}{\urlprefix }}%
\providecommand \urlprefix  [0]{URL }%
\providecommand \Eprint [0]{\href }%
\providecommand \doibase [0]{http://dx.doi.org/}%
\providecommand \selectlanguage [0]{\@gobble}%
\providecommand \bibinfo  [0]{\@secondoftwo}%
\providecommand \bibfield  [0]{\@secondoftwo}%
\providecommand \translation [1]{[#1]}%
\providecommand \BibitemOpen [0]{}%
\providecommand \bibitemStop [0]{}%
\providecommand \bibitemNoStop [0]{.\EOS\space}%
\providecommand \EOS [0]{\spacefactor3000\relax}%
\providecommand \BibitemShut  [1]{\csname bibitem#1\endcsname}%
\let\auto@bib@innerbib\@empty
\bibitem [{\citenamefont {Wollan}\ and\ \citenamefont
  {Koehler}(1955)}]{wollan1955neutron}%
  \BibitemOpen
  \bibfield  {author} {\bibinfo {author} {\bibfnamefont {E.~O.}\ \bibnamefont
  {Wollan}}\ and\ \bibinfo {author} {\bibfnamefont {W.~C.}\ \bibnamefont
  {Koehler}},\ }\href@noop {} {\bibfield  {journal} {\bibinfo  {journal} {Phys.
  Rev.}\ }\textbf {\bibinfo {volume} {100}},\ \bibinfo {pages} {545} (\bibinfo
  {year} {1955})}\BibitemShut {NoStop}%
\bibitem [{\citenamefont {Goodenough}(1955)}]{goodenough1955theory}%
  \BibitemOpen
  \bibfield  {author} {\bibinfo {author} {\bibfnamefont {J.~B.}\ \bibnamefont
  {Goodenough}},\ }\href@noop {} {\bibfield  {journal} {\bibinfo  {journal}
  {Phys. Rev.}\ }\textbf {\bibinfo {volume} {100}},\ \bibinfo {pages} {564}
  (\bibinfo {year} {1955})}\BibitemShut {NoStop}%
\bibitem [{\citenamefont {Radaelli}\ \emph {et~al.}(1997)\citenamefont
  {Radaelli}, \citenamefont {Cox}, \citenamefont {Marezio},\ and\ \citenamefont
  {Cheong}}]{radaelli1997charge}%
  \BibitemOpen
  \bibfield  {author} {\bibinfo {author} {\bibfnamefont {P.~G.}\ \bibnamefont
  {Radaelli}}, \bibinfo {author} {\bibfnamefont {D.~E.}\ \bibnamefont {Cox}},
  \bibinfo {author} {\bibfnamefont {M.}~\bibnamefont {Marezio}}, \ and\
  \bibinfo {author} {\bibfnamefont {S.~W.}\ \bibnamefont {Cheong}},\
  }\href@noop {} {\bibfield  {journal} {\bibinfo  {journal} {Phys. Rev. B}\
  }\textbf {\bibinfo {volume} {55}},\ \bibinfo {pages} {3015} (\bibinfo {year}
  {1997})}\BibitemShut {NoStop}%
\bibitem [{\citenamefont {Przenios{\l}o}\ \emph {et~al.}(2002)\citenamefont
  {Przenios{\l}o}, \citenamefont {Sosnowska}, \citenamefont {Suard},
  \citenamefont {Hewat},\ and\ \citenamefont {Fitch}}]{przenioslo2002phase}%
  \BibitemOpen
  \bibfield  {author} {\bibinfo {author} {\bibfnamefont {R.}~\bibnamefont
  {Przenios{\l}o}}, \bibinfo {author} {\bibfnamefont {I.}~\bibnamefont
  {Sosnowska}}, \bibinfo {author} {\bibfnamefont {E.}~\bibnamefont {Suard}},
  \bibinfo {author} {\bibfnamefont {A.}~\bibnamefont {Hewat}}, \ and\ \bibinfo
  {author} {\bibfnamefont {A.~N.}\ \bibnamefont {Fitch}},\ }\href@noop {}
  {\bibfield  {journal} {\bibinfo  {journal} {J. Phys.: Condens. Matter}\
  }\textbf {\bibinfo {volume} {14}},\ \bibinfo {pages} {5747} (\bibinfo {year}
  {2002})}\BibitemShut {NoStop}%
\bibitem [{\citenamefont {Glazkova}\ \emph {et~al.}(2015)\citenamefont
  {Glazkova}, \citenamefont {Terada}, \citenamefont {Matsushita}, \citenamefont
  {Katsuya}, \citenamefont {Tanaka}, \citenamefont {Sobolev}, \citenamefont
  {Presniakov},\ and\ \citenamefont {Belik}}]{glazkova2015high}%
  \BibitemOpen
  \bibfield  {author} {\bibinfo {author} {\bibfnamefont {Y.~S.}\ \bibnamefont
  {Glazkova}}, \bibinfo {author} {\bibfnamefont {N.}~\bibnamefont {Terada}},
  \bibinfo {author} {\bibfnamefont {Y.}~\bibnamefont {Matsushita}}, \bibinfo
  {author} {\bibfnamefont {Y.}~\bibnamefont {Katsuya}}, \bibinfo {author}
  {\bibfnamefont {M.}~\bibnamefont {Tanaka}}, \bibinfo {author} {\bibfnamefont
  {A.~V.}\ \bibnamefont {Sobolev}}, \bibinfo {author} {\bibfnamefont {I.~A.}\
  \bibnamefont {Presniakov}}, \ and\ \bibinfo {author} {\bibfnamefont {A.~A.}\
  \bibnamefont {Belik}},\ }\href@noop {} {\bibfield  {journal} {\bibinfo
  {journal} {Inorg. Chem.}\ }\textbf {\bibinfo {volume} {54}},\ \bibinfo
  {pages} {9081} (\bibinfo {year} {2015})}\BibitemShut {NoStop}%
\bibitem [{\citenamefont {Locherer}\ \emph {et~al.}(2012)\citenamefont
  {Locherer}, \citenamefont {Dinnebier}, \citenamefont {Kremer}, \citenamefont
  {Greenblatt},\ and\ \citenamefont {Jansen}}]{locherer2012synthesis}%
  \BibitemOpen
  \bibfield  {author} {\bibinfo {author} {\bibfnamefont {T.}~\bibnamefont
  {Locherer}}, \bibinfo {author} {\bibfnamefont {R.}~\bibnamefont {Dinnebier}},
  \bibinfo {author} {\bibfnamefont {R.~K.}\ \bibnamefont {Kremer}}, \bibinfo
  {author} {\bibfnamefont {M.}~\bibnamefont {Greenblatt}}, \ and\ \bibinfo
  {author} {\bibfnamefont {M.}~\bibnamefont {Jansen}},\ }\href@noop {}
  {\bibfield  {journal} {\bibinfo  {journal} {J. Solid State Chem.}\ }\textbf
  {\bibinfo {volume} {190}},\ \bibinfo {pages} {277} (\bibinfo {year}
  {2012})}\BibitemShut {NoStop}%
\bibitem [{\citenamefont {Khomskii}\ and\ \citenamefont {van~den
  Brink}(2000)}]{khomskii2000anharmonic}%
  \BibitemOpen
  \bibfield  {author} {\bibinfo {author} {\bibfnamefont {D.}~\bibnamefont
  {Khomskii}}\ and\ \bibinfo {author} {\bibfnamefont {J.}~\bibnamefont {van~den
  Brink}},\ }\href@noop {} {\bibfield  {journal} {\bibinfo  {journal} {Phys.
  Rev. Lett.}\ }\textbf {\bibinfo {volume} {85}},\ \bibinfo {pages} {3329}
  (\bibinfo {year} {2000})}\BibitemShut {NoStop}%
\bibitem [{\citenamefont {Perks}\ \emph {et~al.}(2012)\citenamefont {Perks},
  \citenamefont {Johnson}, \citenamefont {Martin}, \citenamefont {Chapon},\
  and\ \citenamefont {Radaelli}}]{perks2012magneto}%
  \BibitemOpen
  \bibfield  {author} {\bibinfo {author} {\bibfnamefont {N.~J.}\ \bibnamefont
  {Perks}}, \bibinfo {author} {\bibfnamefont {R.~D.}\ \bibnamefont {Johnson}},
  \bibinfo {author} {\bibfnamefont {C.}~\bibnamefont {Martin}}, \bibinfo
  {author} {\bibfnamefont {L.~C.}\ \bibnamefont {Chapon}}, \ and\ \bibinfo
  {author} {\bibfnamefont {P.~G.}\ \bibnamefont {Radaelli}},\ }\href@noop {}
  {\bibfield  {journal} {\bibinfo  {journal} {Nat. Commun}\ }\textbf {\bibinfo
  {volume} {3}},\ \bibinfo {pages} {1277} (\bibinfo {year} {2012})}\BibitemShut
  {NoStop}%
\bibitem [{\citenamefont {Belik}\ \emph {et~al.}(2016)\citenamefont {Belik},
  \citenamefont {Glazkova}, \citenamefont {Katsuya}, \citenamefont {Tanaka},
  \citenamefont {Sobolev},\ and\ \citenamefont {Presniakov}}]{belik2016low}%
  \BibitemOpen
  \bibfield  {author} {\bibinfo {author} {\bibfnamefont {A.~A.}\ \bibnamefont
  {Belik}}, \bibinfo {author} {\bibfnamefont {Y.~S.}\ \bibnamefont {Glazkova}},
  \bibinfo {author} {\bibfnamefont {Y.}~\bibnamefont {Katsuya}}, \bibinfo
  {author} {\bibfnamefont {M.}~\bibnamefont {Tanaka}}, \bibinfo {author}
  {\bibfnamefont {A.~V.}\ \bibnamefont {Sobolev}}, \ and\ \bibinfo {author}
  {\bibfnamefont {I.~A.}\ \bibnamefont {Presniakov}},\ }\href@noop {}
  {\bibfield  {journal} {\bibinfo  {journal} {J. Phys. Chem. C}\ }\textbf
  {\bibinfo {volume} {120}},\ \bibinfo {pages} {8278} (\bibinfo {year}
  {2016})}\BibitemShut {NoStop}%
\bibitem [{\citenamefont {Johnson}\ \emph {et~al.}(2017)\citenamefont
  {Johnson}, \citenamefont {Khalyavin}, \citenamefont {Manuel}, \citenamefont
  {Radaelli}, \citenamefont {Glazkova}, \citenamefont {Terada},\ and\
  \citenamefont {Belik}}]{johnson2017magneto}%
  \BibitemOpen
  \bibfield  {author} {\bibinfo {author} {\bibfnamefont {R.~D.}\ \bibnamefont
  {Johnson}}, \bibinfo {author} {\bibfnamefont {D.~D.}\ \bibnamefont
  {Khalyavin}}, \bibinfo {author} {\bibfnamefont {P.}~\bibnamefont {Manuel}},
  \bibinfo {author} {\bibfnamefont {P.~G.}\ \bibnamefont {Radaelli}}, \bibinfo
  {author} {\bibfnamefont {I.~S.}\ \bibnamefont {Glazkova}}, \bibinfo {author}
  {\bibfnamefont {N.}~\bibnamefont {Terada}}, \ and\ \bibinfo {author}
  {\bibfnamefont {A.~A.}\ \bibnamefont {Belik}},\ }\href@noop {} {\bibfield
  {journal} {\bibinfo  {journal} {Phys. Rev. B}\ }\textbf {\bibinfo {volume}
  {96}},\ \bibinfo {pages} {054448} (\bibinfo {year} {2017})}\BibitemShut
  {NoStop}%
\bibitem [{\citenamefont {Guo}\ \emph {et~al.}(2017)\citenamefont {Guo},
  \citenamefont {Fern{\'a}ndez-D{\'\i}az}, \citenamefont {Zhou}, \citenamefont
  {Yin}, \citenamefont {Long},\ and\ \citenamefont {Komarek}}]{guo2017non}%
  \BibitemOpen
  \bibfield  {author} {\bibinfo {author} {\bibfnamefont {H.}~\bibnamefont
  {Guo}}, \bibinfo {author} {\bibfnamefont {M.}~\bibnamefont
  {Fern{\'a}ndez-D{\'\i}az}}, \bibinfo {author} {\bibfnamefont
  {L.}~\bibnamefont {Zhou}}, \bibinfo {author} {\bibfnamefont {Y.}~\bibnamefont
  {Yin}}, \bibinfo {author} {\bibfnamefont {Y.}~\bibnamefont {Long}}, \ and\
  \bibinfo {author} {\bibfnamefont {A.~C.}\ \bibnamefont {Komarek}},\
  }\href@noop {} {\bibfield  {journal} {\bibinfo  {journal} {Sci. Rep.}\
  }\textbf {\bibinfo {volume} {7}},\ \bibinfo {pages} {45939} (\bibinfo {year}
  {2017})}\BibitemShut {NoStop}%
\bibitem [{\citenamefont {Johnson}\ \emph {et~al.}(2021)\citenamefont
  {Johnson}, \citenamefont {Khalyavin}, \citenamefont {Manuel},\ and\
  \citenamefont {Belik}}]{johnson2021competing}%
  \BibitemOpen
  \bibfield  {author} {\bibinfo {author} {\bibfnamefont {R.~D.}\ \bibnamefont
  {Johnson}}, \bibinfo {author} {\bibfnamefont {D.~D.}\ \bibnamefont
  {Khalyavin}}, \bibinfo {author} {\bibfnamefont {P.}~\bibnamefont {Manuel}}, \
  and\ \bibinfo {author} {\bibfnamefont {A.~A.}\ \bibnamefont {Belik}},\
  }\href@noop {} {\bibfield  {journal} {\bibinfo  {journal} {Phys. Rev. B}\
  }\textbf {\bibinfo {volume} {103}},\ \bibinfo {pages} {134115} (\bibinfo
  {year} {2021})}\BibitemShut {NoStop}%
\bibitem [{\citenamefont {Chen}\ \emph {et~al.}(2021)\citenamefont {Chen},
  \citenamefont {Wang}, \citenamefont {Cheng}, \citenamefont {Chuang},
  \citenamefont {Simonov}, \citenamefont {Bristowe},\ and\ \citenamefont
  {Senn}}]{chen2021striping}%
  \BibitemOpen
  \bibfield  {author} {\bibinfo {author} {\bibfnamefont {W.~T.}\ \bibnamefont
  {Chen}}, \bibinfo {author} {\bibfnamefont {C.~W.}\ \bibnamefont {Wang}},
  \bibinfo {author} {\bibfnamefont {C.~C.}\ \bibnamefont {Cheng}}, \bibinfo
  {author} {\bibfnamefont {Y.~C.}\ \bibnamefont {Chuang}}, \bibinfo {author}
  {\bibfnamefont {A.}~\bibnamefont {Simonov}}, \bibinfo {author} {\bibfnamefont
  {N.~C.}\ \bibnamefont {Bristowe}}, \ and\ \bibinfo {author} {\bibfnamefont
  {M.~S.}\ \bibnamefont {Senn}},\ }\href@noop {} {\bibfield  {journal}
  {\bibinfo  {journal} {Nat. Commun.}\ }\textbf {\bibinfo {volume} {12}},\
  \bibinfo {pages} {6319} (\bibinfo {year} {2021})}\BibitemShut {NoStop}%
\bibitem [{\citenamefont {Radaelli}\ \emph {et~al.}(1995)\citenamefont
  {Radaelli}, \citenamefont {Cox}, \citenamefont {Marezio}, \citenamefont
  {Cheong}, \citenamefont {Schiffer},\ and\ \citenamefont
  {Ramirez}}]{radaelli1995simultaneous}%
  \BibitemOpen
  \bibfield  {author} {\bibinfo {author} {\bibfnamefont {P.~G.}\ \bibnamefont
  {Radaelli}}, \bibinfo {author} {\bibfnamefont {D.~E.}\ \bibnamefont {Cox}},
  \bibinfo {author} {\bibfnamefont {M.}~\bibnamefont {Marezio}}, \bibinfo
  {author} {\bibfnamefont {S.-W.}\ \bibnamefont {Cheong}}, \bibinfo {author}
  {\bibfnamefont {P.~E.}\ \bibnamefont {Schiffer}}, \ and\ \bibinfo {author}
  {\bibfnamefont {A.~P.}\ \bibnamefont {Ramirez}},\ }\href@noop {} {\bibfield
  {journal} {\bibinfo  {journal} {Phys. Rev. Lett.}\ }\textbf {\bibinfo
  {volume} {75}},\ \bibinfo {pages} {4488} (\bibinfo {year}
  {1995})}\BibitemShut {NoStop}%
\bibitem [{\citenamefont {Hwang}\ \emph
  {et~al.}(1995{\natexlab{a}})\citenamefont {Hwang}, \citenamefont {Cheong},
  \citenamefont {Radaelli}, \citenamefont {Marezio},\ and\ \citenamefont
  {Batlogg}}]{hwang1995lattice}%
  \BibitemOpen
  \bibfield  {author} {\bibinfo {author} {\bibfnamefont {H.~Y.}\ \bibnamefont
  {Hwang}}, \bibinfo {author} {\bibfnamefont {S.-W.}\ \bibnamefont {Cheong}},
  \bibinfo {author} {\bibfnamefont {P.~G.}\ \bibnamefont {Radaelli}}, \bibinfo
  {author} {\bibfnamefont {M.}~\bibnamefont {Marezio}}, \ and\ \bibinfo
  {author} {\bibfnamefont {B.}~\bibnamefont {Batlogg}},\ }\href@noop {}
  {\bibfield  {journal} {\bibinfo  {journal} {Phys. Rev. Lett.}\ }\textbf
  {\bibinfo {volume} {75}},\ \bibinfo {pages} {914} (\bibinfo {year}
  {1995}{\natexlab{a}})}\BibitemShut {NoStop}%
\bibitem [{\citenamefont {Rodriguez-Carvajal}\ \emph
  {et~al.}(1998)\citenamefont {Rodriguez-Carvajal}, \citenamefont {Hennion},
  \citenamefont {Moussa}, \citenamefont {Moudden}, \citenamefont {Pinsard},\
  and\ \citenamefont {Revcolevschi}}]{rodriguez1998neutron}%
  \BibitemOpen
  \bibfield  {author} {\bibinfo {author} {\bibfnamefont {J.}~\bibnamefont
  {Rodriguez-Carvajal}}, \bibinfo {author} {\bibfnamefont {M.}~\bibnamefont
  {Hennion}}, \bibinfo {author} {\bibfnamefont {F.}~\bibnamefont {Moussa}},
  \bibinfo {author} {\bibfnamefont {A.~H.}\ \bibnamefont {Moudden}}, \bibinfo
  {author} {\bibfnamefont {L.}~\bibnamefont {Pinsard}}, \ and\ \bibinfo
  {author} {\bibfnamefont {A.}~\bibnamefont {Revcolevschi}},\ }\href@noop {}
  {\bibfield  {journal} {\bibinfo  {journal} {Phys. Rev. B}\ }\textbf {\bibinfo
  {volume} {57}},\ \bibinfo {pages} {R3189} (\bibinfo {year}
  {1998})}\BibitemShut {NoStop}%
\bibitem [{\citenamefont {Van~Aken}\ \emph {et~al.}(2003)\citenamefont
  {Van~Aken}, \citenamefont {Jurchescu}, \citenamefont {Meetsma}, \citenamefont
  {Tomioka}, \citenamefont {Tokura},\ and\ \citenamefont
  {Palstra}}]{van2003orbital}%
  \BibitemOpen
  \bibfield  {author} {\bibinfo {author} {\bibfnamefont {B.~B.}\ \bibnamefont
  {Van~Aken}}, \bibinfo {author} {\bibfnamefont {O.~D.}\ \bibnamefont
  {Jurchescu}}, \bibinfo {author} {\bibfnamefont {A.}~\bibnamefont {Meetsma}},
  \bibinfo {author} {\bibfnamefont {Y.}~\bibnamefont {Tomioka}}, \bibinfo
  {author} {\bibfnamefont {Y.}~\bibnamefont {Tokura}}, \ and\ \bibinfo {author}
  {\bibfnamefont {T.~T.~M.}\ \bibnamefont {Palstra}},\ }\href@noop {}
  {\bibfield  {journal} {\bibinfo  {journal} {Phys. Rev. Lett.}\ }\textbf
  {\bibinfo {volume} {90}},\ \bibinfo {pages} {066403} (\bibinfo {year}
  {2003})}\BibitemShut {NoStop}%
\bibitem [{\citenamefont {Cheong}\ and\ \citenamefont
  {Hwang}(2000)}]{cheong2000colossal}%
  \BibitemOpen
  \bibfield  {author} {\bibinfo {author} {\bibfnamefont {S.~W.}\ \bibnamefont
  {Cheong}}\ and\ \bibinfo {author} {\bibfnamefont {H.~Y.}\ \bibnamefont
  {Hwang}},\ }in\ \href@noop {} {\emph {\bibinfo {booktitle} {Colossal
  Magnetoresistance Oxides}}},\ \bibinfo {editor} {edited by\ \bibinfo {editor}
  {\bibfnamefont {Y.}~\bibnamefont {Tokura}}}\ (\bibinfo  {publisher} {Gordon
  and Breach},\ \bibinfo {address} {London},\ \bibinfo {year} {2000})\
  Chap.~\bibinfo {chapter} {7}\BibitemShut {NoStop}%
\bibitem [{\citenamefont {Hwang}\ \emph
  {et~al.}(1995{\natexlab{b}})\citenamefont {Hwang}, \citenamefont {Palstra},
  \citenamefont {Cheong},\ and\ \citenamefont {Batlogg}}]{hwang1995pressure}%
  \BibitemOpen
  \bibfield  {author} {\bibinfo {author} {\bibfnamefont {H.~Y.}\ \bibnamefont
  {Hwang}}, \bibinfo {author} {\bibfnamefont {T.~T.~M.}\ \bibnamefont
  {Palstra}}, \bibinfo {author} {\bibfnamefont {S.~W.}\ \bibnamefont {Cheong}},
  \ and\ \bibinfo {author} {\bibfnamefont {B.}~\bibnamefont {Batlogg}},\
  }\href@noop {} {\bibfield  {journal} {\bibinfo  {journal} {Phys. Rev. B}\
  }\textbf {\bibinfo {volume} {52}},\ \bibinfo {pages} {15046} (\bibinfo {year}
  {1995}{\natexlab{b}})}\BibitemShut {NoStop}%
\bibitem [{\citenamefont {Chen}\ \emph {et~al.}(2018)\citenamefont {Chen},
  \citenamefont {Wang}, \citenamefont {Wu}, \citenamefont {Chou}, \citenamefont
  {Yang}, \citenamefont {Simonov},\ and\ \citenamefont
  {Senn}}]{chen2018improper}%
  \BibitemOpen
  \bibfield  {author} {\bibinfo {author} {\bibfnamefont {W.-T.}\ \bibnamefont
  {Chen}}, \bibinfo {author} {\bibfnamefont {C.-W.}\ \bibnamefont {Wang}},
  \bibinfo {author} {\bibfnamefont {H.-C.}\ \bibnamefont {Wu}}, \bibinfo
  {author} {\bibfnamefont {F.-C.}\ \bibnamefont {Chou}}, \bibinfo {author}
  {\bibfnamefont {H.-D.}\ \bibnamefont {Yang}}, \bibinfo {author}
  {\bibfnamefont {A.}~\bibnamefont {Simonov}}, \ and\ \bibinfo {author}
  {\bibfnamefont {M.~S.}\ \bibnamefont {Senn}},\ }\href@noop {} {\bibfield
  {journal} {\bibinfo  {journal} {Phys. Rev. B}\ }\textbf {\bibinfo {volume}
  {97}},\ \bibinfo {pages} {144102} (\bibinfo {year} {2018})}\BibitemShut
  {NoStop}%
\bibitem [{\citenamefont {Belik}\ \emph {et~al.}(2017)\citenamefont {Belik},
  \citenamefont {Matsushita}, \citenamefont {Kumagai}, \citenamefont {Katsuya},
  \citenamefont {Tanaka}, \citenamefont {Stefanovich}, \citenamefont
  {Lazoryak}, \citenamefont {Oba},\ and\ \citenamefont
  {Yamaura}}]{belik2017complex}%
  \BibitemOpen
  \bibfield  {author} {\bibinfo {author} {\bibfnamefont {A.~A.}\ \bibnamefont
  {Belik}}, \bibinfo {author} {\bibfnamefont {Y.}~\bibnamefont {Matsushita}},
  \bibinfo {author} {\bibfnamefont {Y.}~\bibnamefont {Kumagai}}, \bibinfo
  {author} {\bibfnamefont {Y.}~\bibnamefont {Katsuya}}, \bibinfo {author}
  {\bibfnamefont {M.}~\bibnamefont {Tanaka}}, \bibinfo {author} {\bibfnamefont
  {S.~Y.}\ \bibnamefont {Stefanovich}}, \bibinfo {author} {\bibfnamefont
  {B.~I.}\ \bibnamefont {Lazoryak}}, \bibinfo {author} {\bibfnamefont
  {F.}~\bibnamefont {Oba}}, \ and\ \bibinfo {author} {\bibfnamefont
  {K.}~\bibnamefont {Yamaura}},\ }\href@noop {} {\bibfield  {journal} {\bibinfo
   {journal} {Inorg. Chem.}\ }\textbf {\bibinfo {volume} {56}},\ \bibinfo
  {pages} {12272} (\bibinfo {year} {2017})}\BibitemShut {NoStop}%
\bibitem [{\citenamefont {Khalyavin}\ \emph {et~al.}(2020)\citenamefont
  {Khalyavin}, \citenamefont {Johnson}, \citenamefont {Orlandi}, \citenamefont
  {Radaelli}, \citenamefont {Manuel},\ and\ \citenamefont
  {Belik}}]{khalyavin2020emergent}%
  \BibitemOpen
  \bibfield  {author} {\bibinfo {author} {\bibfnamefont {D.~D.}\ \bibnamefont
  {Khalyavin}}, \bibinfo {author} {\bibfnamefont {R.~D.}\ \bibnamefont
  {Johnson}}, \bibinfo {author} {\bibfnamefont {F.}~\bibnamefont {Orlandi}},
  \bibinfo {author} {\bibfnamefont {P.~G.}\ \bibnamefont {Radaelli}}, \bibinfo
  {author} {\bibfnamefont {P.}~\bibnamefont {Manuel}}, \ and\ \bibinfo {author}
  {\bibfnamefont {A.~A.}\ \bibnamefont {Belik}},\ }\href@noop {} {\bibfield
  {journal} {\bibinfo  {journal} {Science}\ }\textbf {\bibinfo {volume}
  {369}},\ \bibinfo {pages} {680} (\bibinfo {year} {2020})}\BibitemShut
  {NoStop}%
\bibitem [{\citenamefont {Campbell}\ \emph {et~al.}(2006)\citenamefont
  {Campbell}, \citenamefont {Stokes}, \citenamefont {Tanner},\ and\
  \citenamefont {Hatch}}]{Campbell1}%
  \BibitemOpen
  \bibfield  {author} {\bibinfo {author} {\bibfnamefont {B.~J.}\ \bibnamefont
  {Campbell}}, \bibinfo {author} {\bibfnamefont {H.~T.}\ \bibnamefont
  {Stokes}}, \bibinfo {author} {\bibfnamefont {D.~E.}\ \bibnamefont {Tanner}},
  \ and\ \bibinfo {author} {\bibfnamefont {D.~M.}\ \bibnamefont {Hatch}},\
  }\href@noop {} {\bibfield  {journal} {\bibinfo  {journal} {J. Appl.
  Crystallogr.}\ }\textbf {\bibinfo {volume} {39}},\ \bibinfo {pages} {607}
  (\bibinfo {year} {2006})}\BibitemShut {NoStop}%
\bibitem [{\citenamefont {Senn}\ and\ \citenamefont
  {Bristowe}(2018)}]{senn2018group}%
  \BibitemOpen
  \bibfield  {author} {\bibinfo {author} {\bibfnamefont {M.~S.}\ \bibnamefont
  {Senn}}\ and\ \bibinfo {author} {\bibfnamefont {N.~C.}\ \bibnamefont
  {Bristowe}},\ }\href@noop {} {\bibfield  {journal} {\bibinfo  {journal} {Acta
  Crystallogr., Sect. A: Found. Adv.}\ }\textbf {\bibinfo {volume} {74}},\
  \bibinfo {pages} {308} (\bibinfo {year} {2018})}\BibitemShut {NoStop}%
\bibitem [{\citenamefont {Streltsov}\ and\ \citenamefont
  {Khomskii}(2014)}]{streltsov2014jahn}%
  \BibitemOpen
  \bibfield  {author} {\bibinfo {author} {\bibfnamefont {S.~V.}\ \bibnamefont
  {Streltsov}}\ and\ \bibinfo {author} {\bibfnamefont {D.~I.}\ \bibnamefont
  {Khomskii}},\ }\href@noop {} {\bibfield  {journal} {\bibinfo  {journal}
  {Phys. Rev. B}\ }\textbf {\bibinfo {volume} {89}},\ \bibinfo {pages} {201115}
  (\bibinfo {year} {2014})}\BibitemShut {NoStop}%
\bibitem [{\citenamefont {Shannon}(1976)}]{shannon1976revised}%
  \BibitemOpen
  \bibfield  {author} {\bibinfo {author} {\bibfnamefont {R.~D.}\ \bibnamefont
  {Shannon}},\ }\href@noop {} {\bibfield  {journal} {\bibinfo  {journal} {Acta
  Crystallogr., Sect. A: Cryst. Phys., Diffr., Theor. Gen. Crystallogr.}\
  }\textbf {\bibinfo {volume} {32}},\ \bibinfo {pages} {751} (\bibinfo {year}
  {1976})}\BibitemShut {NoStop}%
\bibitem [{\citenamefont {Glazer}(1972)}]{glazer1972classification}%
  \BibitemOpen
  \bibfield  {author} {\bibinfo {author} {\bibfnamefont {A.~M.}\ \bibnamefont
  {Glazer}},\ }\href@noop {} {\bibfield  {journal} {\bibinfo  {journal} {Acta
  Crystallogr., Sect. B: Struct. Crystallogr. Cryst. Chem.}\ }\textbf {\bibinfo
  {volume} {28}},\ \bibinfo {pages} {3384} (\bibinfo {year}
  {1972})}\BibitemShut {NoStop}%
\bibitem [{\citenamefont {Tidey}\ \emph {et~al.}(2022)\citenamefont {Tidey},
  \citenamefont {Liu}, \citenamefont {Lai}, \citenamefont {Chuang},
  \citenamefont {Chen}, \citenamefont {Cane}, \citenamefont {Lester},
  \citenamefont {Petsch}, \citenamefont {Herlihy}, \citenamefont {Simonov},
  \citenamefont {Hayden},\ and\ \citenamefont {Senn}}]{tidey2022pronounced}%
  \BibitemOpen
  \bibfield  {author} {\bibinfo {author} {\bibfnamefont {J.~P.}\ \bibnamefont
  {Tidey}}, \bibinfo {author} {\bibfnamefont {E.-P.}\ \bibnamefont {Liu}},
  \bibinfo {author} {\bibfnamefont {Y.-C.}\ \bibnamefont {Lai}}, \bibinfo
  {author} {\bibfnamefont {Y.-C.}\ \bibnamefont {Chuang}}, \bibinfo {author}
  {\bibfnamefont {W.-T.}\ \bibnamefont {Chen}}, \bibinfo {author}
  {\bibfnamefont {L.~J.}\ \bibnamefont {Cane}}, \bibinfo {author}
  {\bibfnamefont {C.}~\bibnamefont {Lester}}, \bibinfo {author} {\bibfnamefont
  {A.~N.}\ \bibnamefont {Petsch}}, \bibinfo {author} {\bibfnamefont
  {A.}~\bibnamefont {Herlihy}}, \bibinfo {author} {\bibfnamefont
  {A.}~\bibnamefont {Simonov}}, \bibinfo {author} {\bibfnamefont {S.~M.}\
  \bibnamefont {Hayden}}, \ and\ \bibinfo {author} {\bibfnamefont {M.~S.}\
  \bibnamefont {Senn}},\ }\href@noop {} {\bibfield  {journal} {\bibinfo
  {journal} {Sci. Rep.}\ }\textbf {\bibinfo {volume} {12}},\ \bibinfo {pages}
  {14343} (\bibinfo {year} {2022})}\BibitemShut {NoStop}%
\bibitem [{\citenamefont {Uehara}\ \emph {et~al.}(1999)\citenamefont {Uehara},
  \citenamefont {Mori}, \citenamefont {Chen},\ and\ \citenamefont
  {Cheong}}]{uehara1999percolative}%
  \BibitemOpen
  \bibfield  {author} {\bibinfo {author} {\bibfnamefont {M.}~\bibnamefont
  {Uehara}}, \bibinfo {author} {\bibfnamefont {S.}~\bibnamefont {Mori}},
  \bibinfo {author} {\bibfnamefont {C.}~\bibnamefont {Chen}}, \ and\ \bibinfo
  {author} {\bibfnamefont {S.-W.}\ \bibnamefont {Cheong}},\ }\href@noop {}
  {\bibfield  {journal} {\bibinfo  {journal} {Nature}\ }\textbf {\bibinfo
  {volume} {399}},\ \bibinfo {pages} {560} (\bibinfo {year}
  {1999})}\BibitemShut {NoStop}%
\bibitem [{\citenamefont {Sage}\ \emph {et~al.}(2006)\citenamefont {Sage},
  \citenamefont {Blake}, \citenamefont {Nieuwenhuys},\ and\ \citenamefont
  {Palstra}}]{sage2006evidence}%
  \BibitemOpen
  \bibfield  {author} {\bibinfo {author} {\bibfnamefont {M.~H.}\ \bibnamefont
  {Sage}}, \bibinfo {author} {\bibfnamefont {G.~R.}\ \bibnamefont {Blake}},
  \bibinfo {author} {\bibfnamefont {G.~J.}\ \bibnamefont {Nieuwenhuys}}, \ and\
  \bibinfo {author} {\bibfnamefont {T.~T.~M.}\ \bibnamefont {Palstra}},\
  }\href@noop {} {\bibfield  {journal} {\bibinfo  {journal} {Phys. Rev. Lett.}\
  }\textbf {\bibinfo {volume} {96}},\ \bibinfo {pages} {036401} (\bibinfo
  {year} {2006})}\BibitemShut {NoStop}%
\bibitem [{\citenamefont {Ritter}\ \emph {et~al.}(2016)\citenamefont {Ritter},
  \citenamefont {Bazuev},\ and\ \citenamefont
  {Fauth}}]{ritter2016crystallographic}%
  \BibitemOpen
  \bibfield  {author} {\bibinfo {author} {\bibfnamefont {C.}~\bibnamefont
  {Ritter}}, \bibinfo {author} {\bibfnamefont {G.~V.}\ \bibnamefont {Bazuev}},
  \ and\ \bibinfo {author} {\bibfnamefont {F.}~\bibnamefont {Fauth}},\
  }\href@noop {} {\bibfield  {journal} {\bibinfo  {journal} {Phys. Rev. B}\
  }\textbf {\bibinfo {volume} {93}},\ \bibinfo {pages} {054423} (\bibinfo
  {year} {2016})}\BibitemShut {NoStop}%
\bibitem [{\citenamefont {Saha}\ \emph {et~al.}(2017)\citenamefont {Saha},
  \citenamefont {Fauth}, \citenamefont {Caignaert},\ and\ \citenamefont
  {Sundaresan}}]{saha2017coexistence}%
  \BibitemOpen
  \bibfield  {author} {\bibinfo {author} {\bibfnamefont {R.}~\bibnamefont
  {Saha}}, \bibinfo {author} {\bibfnamefont {F.}~\bibnamefont {Fauth}},
  \bibinfo {author} {\bibfnamefont {V.}~\bibnamefont {Caignaert}}, \ and\
  \bibinfo {author} {\bibfnamefont {A.}~\bibnamefont {Sundaresan}},\
  }\href@noop {} {\bibfield  {journal} {\bibinfo  {journal} {Phys. Rev. B}\
  }\textbf {\bibinfo {volume} {95}},\ \bibinfo {pages} {184107} (\bibinfo
  {year} {2017})}\BibitemShut {NoStop}%
\bibitem [{\citenamefont {Margadonna}\ and\ \citenamefont
  {Karotsis}(2006)}]{margadonna2006cooperative}%
  \BibitemOpen
  \bibfield  {author} {\bibinfo {author} {\bibfnamefont {S.}~\bibnamefont
  {Margadonna}}\ and\ \bibinfo {author} {\bibfnamefont {G.}~\bibnamefont
  {Karotsis}},\ }\href@noop {} {\bibfield  {journal} {\bibinfo  {journal} {J.
  Am. Chem. Soc.}\ }\textbf {\bibinfo {volume} {128}},\ \bibinfo {pages}
  {16436} (\bibinfo {year} {2006})}\BibitemShut {NoStop}%
\bibitem [{\citenamefont {Zhou}\ \emph {et~al.}(2011)\citenamefont {Zhou},
  \citenamefont {Alonso}, \citenamefont {Han}, \citenamefont
  {Fern{\'a}ndez-D{\'\i}az}, \citenamefont {Cheng},\ and\ \citenamefont
  {Goodenough}}]{zhou2011jahn}%
  \BibitemOpen
  \bibfield  {author} {\bibinfo {author} {\bibfnamefont {J.-S.}\ \bibnamefont
  {Zhou}}, \bibinfo {author} {\bibfnamefont {J.~A.}\ \bibnamefont {Alonso}},
  \bibinfo {author} {\bibfnamefont {J.~T.}\ \bibnamefont {Han}}, \bibinfo
  {author} {\bibfnamefont {M.~T.}\ \bibnamefont {Fern{\'a}ndez-D{\'\i}az}},
  \bibinfo {author} {\bibfnamefont {J.-G.}\ \bibnamefont {Cheng}}, \ and\
  \bibinfo {author} {\bibfnamefont {J.~B.}\ \bibnamefont {Goodenough}},\
  }\href@noop {} {\bibfield  {journal} {\bibinfo  {journal} {J. Fluorine
  Chem.}\ }\textbf {\bibinfo {volume} {132}},\ \bibinfo {pages} {1117}
  (\bibinfo {year} {2011})}\BibitemShut {NoStop}%
\bibitem [{\citenamefont {Long}\ \emph {et~al.}(2009)\citenamefont {Long},
  \citenamefont {Hayashi}, \citenamefont {Saito}, \citenamefont {Azuma},
  \citenamefont {Muranaka},\ and\ \citenamefont
  {Shimakawa}}]{long2009temperature}%
  \BibitemOpen
  \bibfield  {author} {\bibinfo {author} {\bibfnamefont {Y.}~\bibnamefont
  {Long}}, \bibinfo {author} {\bibfnamefont {N.}~\bibnamefont {Hayashi}},
  \bibinfo {author} {\bibfnamefont {T.}~\bibnamefont {Saito}}, \bibinfo
  {author} {\bibfnamefont {M.}~\bibnamefont {Azuma}}, \bibinfo {author}
  {\bibfnamefont {S.}~\bibnamefont {Muranaka}}, \ and\ \bibinfo {author}
  {\bibfnamefont {Y.}~\bibnamefont {Shimakawa}},\ }\href@noop {} {\bibfield
  {journal} {\bibinfo  {journal} {Nature}\ }\textbf {\bibinfo {volume} {458}},\
  \bibinfo {pages} {60} (\bibinfo {year} {2009})}\BibitemShut {NoStop}%
\bibitem [{\citenamefont {Goff}\ and\ \citenamefont
  {Attfield}(2004)}]{goff2004charge}%
  \BibitemOpen
  \bibfield  {author} {\bibinfo {author} {\bibfnamefont {R.~J.}\ \bibnamefont
  {Goff}}\ and\ \bibinfo {author} {\bibfnamefont {J.}~\bibnamefont
  {Attfield}},\ }\href@noop {} {\bibfield  {journal} {\bibinfo  {journal}
  {Phys. Rev. B}\ }\textbf {\bibinfo {volume} {70}},\ \bibinfo {pages} {140404}
  (\bibinfo {year} {2004})}\BibitemShut {NoStop}%
\bibitem [{\citenamefont {Alonso}\ \emph {et~al.}(1999)\citenamefont {Alonso},
  \citenamefont {Garc{\'\i}a-Mu{\~n}oz}, \citenamefont
  {Fern{\'a}ndez-D{\'\i}az}, \citenamefont {Aranda}, \citenamefont
  {Mart{\'\i}nez-Lope},\ and\ \citenamefont {Casais}}]{alonso1999charge}%
  \BibitemOpen
  \bibfield  {author} {\bibinfo {author} {\bibfnamefont {J.~A.}\ \bibnamefont
  {Alonso}}, \bibinfo {author} {\bibfnamefont {J.~L.}\ \bibnamefont
  {Garc{\'\i}a-Mu{\~n}oz}}, \bibinfo {author} {\bibfnamefont {M.~T.}\
  \bibnamefont {Fern{\'a}ndez-D{\'\i}az}}, \bibinfo {author} {\bibfnamefont
  {M.~A.~G.}\ \bibnamefont {Aranda}}, \bibinfo {author} {\bibfnamefont {M.~J.}\
  \bibnamefont {Mart{\'\i}nez-Lope}}, \ and\ \bibinfo {author} {\bibfnamefont
  {M.~T.}\ \bibnamefont {Casais}},\ }\href@noop {} {\bibfield  {journal}
  {\bibinfo  {journal} {Phys. Rev. Lett.}\ }\textbf {\bibinfo {volume} {82}},\
  \bibinfo {pages} {3871} (\bibinfo {year} {1999})}\BibitemShut {NoStop}%
\bibitem [{\citenamefont {Williams}\ and\ \citenamefont
  {Attfield}(2002)}]{williams2002alternative}%
  \BibitemOpen
  \bibfield  {author} {\bibinfo {author} {\bibfnamefont {A.~J.}\ \bibnamefont
  {Williams}}\ and\ \bibinfo {author} {\bibfnamefont {J.~P.}\ \bibnamefont
  {Attfield}},\ }\href@noop {} {\bibfield  {journal} {\bibinfo  {journal}
  {Phys. Rev. B}\ }\textbf {\bibinfo {volume} {66}},\ \bibinfo {pages} {220405}
  (\bibinfo {year} {2002})}\BibitemShut {NoStop}%
\bibitem [{\citenamefont {Tohyama}\ \emph {et~al.}(2013)\citenamefont
  {Tohyama}, \citenamefont {Senn}, \citenamefont {Saito}, \citenamefont {Chen},
  \citenamefont {Tang}, \citenamefont {Attfield},\ and\ \citenamefont
  {Shimakawa}}]{tohyama2013valence}%
  \BibitemOpen
  \bibfield  {author} {\bibinfo {author} {\bibfnamefont {T.}~\bibnamefont
  {Tohyama}}, \bibinfo {author} {\bibfnamefont {M.~S.}\ \bibnamefont {Senn}},
  \bibinfo {author} {\bibfnamefont {T.}~\bibnamefont {Saito}}, \bibinfo
  {author} {\bibfnamefont {W.-t.}\ \bibnamefont {Chen}}, \bibinfo {author}
  {\bibfnamefont {C.~C.}\ \bibnamefont {Tang}}, \bibinfo {author}
  {\bibfnamefont {J.~P.}\ \bibnamefont {Attfield}}, \ and\ \bibinfo {author}
  {\bibfnamefont {Y.}~\bibnamefont {Shimakawa}},\ }\href@noop {} {\bibfield
  {journal} {\bibinfo  {journal} {Chem. Mater.}\ }\textbf {\bibinfo {volume}
  {25}},\ \bibinfo {pages} {178} (\bibinfo {year} {2013})}\BibitemShut
  {NoStop}%
\bibitem [{\citenamefont {Aimi}\ \emph {et~al.}(2014)\citenamefont {Aimi},
  \citenamefont {Mori}, \citenamefont {Hiraki}, \citenamefont {Takahashi},
  \citenamefont {Shan}, \citenamefont {Shirako}, \citenamefont {Zhou},\ and\
  \citenamefont {Inaguma}}]{aimi2014high}%
  \BibitemOpen
  \bibfield  {author} {\bibinfo {author} {\bibfnamefont {A.}~\bibnamefont
  {Aimi}}, \bibinfo {author} {\bibfnamefont {D.}~\bibnamefont {Mori}}, \bibinfo
  {author} {\bibfnamefont {K.-i.}\ \bibnamefont {Hiraki}}, \bibinfo {author}
  {\bibfnamefont {T.}~\bibnamefont {Takahashi}}, \bibinfo {author}
  {\bibfnamefont {Y.~J.}\ \bibnamefont {Shan}}, \bibinfo {author}
  {\bibfnamefont {Y.}~\bibnamefont {Shirako}}, \bibinfo {author} {\bibfnamefont
  {J.}~\bibnamefont {Zhou}}, \ and\ \bibinfo {author} {\bibfnamefont
  {Y.}~\bibnamefont {Inaguma}},\ }\href@noop {} {\bibfield  {journal} {\bibinfo
   {journal} {Chem. Mater.}\ }\textbf {\bibinfo {volume} {26}},\ \bibinfo
  {pages} {2601} (\bibinfo {year} {2014})}\BibitemShut {NoStop}%
\bibitem [{\citenamefont {Hatch}\ and\ \citenamefont {Stokes}(2003)}]{Hatch1}%
  \BibitemOpen
  \bibfield  {author} {\bibinfo {author} {\bibfnamefont {D.~M.}\ \bibnamefont
  {Hatch}}\ and\ \bibinfo {author} {\bibfnamefont {H.~T.}\ \bibnamefont
  {Stokes}},\ }\href@noop {} {\bibfield  {journal} {\bibinfo  {journal} {J.
  Appl. Crystallogr.}\ }\textbf {\bibinfo {volume} {36}},\ \bibinfo {pages}
  {951} (\bibinfo {year} {2003})}\BibitemShut {NoStop}%
\end{thebibliography}%

\end{document}